\documentclass[11pt,a4paper]{article}
\usepackage{jcappub} % for details on the use of the package, please see the JINST-author-manual
\usepackage{lineno}
\usepackage{bm}
\author[a]{Yong Yuan,}
\author[a]{Minghui Du,}
\author[b]{Wen-Fan Feng,}
\author[c,d]{Benyang Zhu,}
\author[a,e]{Qing Diao}
\author[a,f,g]{Peng Xu,}
\author[h,1]{Xilong Fan\note{Corresponding author.}}

\affiliation[a]{Center for Gravitational Wave Experiment, National Microgravity Laboratory, Institute of Mechanics, Chinese Academy of Sciences, Beijing, China}
\affiliation[b]{Kavli Institute for Astronomy and Astrophysics, Peking University, Beijing 100871, China}
\affiliation[c]{Key Laboratory of Dark Matter and Space Astronomy, Purple Mountain Observatory, Chinese Academy of Sciences, 210023, Nanjing, China}
\affiliation[d]{School of Astronomy and Space Science, University of Science and Technology of China, 230026, Hefei, China}
\affiliation[e]{International Centre for Theoretical Physics Asia-Pacific, University of Chinese Academy of Sciences, 100190, Beijing, China}
\affiliation[f]{Taiji Laboratory for Gravitational Wave Universe (Beijing/Hangzhou), University of Chinese Academy of Sciences, Beijing, 100049, China}
\affiliation[g]{Hangzhou Institute for Advanced Study, University of Chinese Academy of Sciences, Hangzhou, 310124, China}
\affiliation[h]{School of Physics Science And Technology, Wuhan University, No.299 Bayi Road, Wuhan, Hubei, China}

\emailAdd{yuanyong@imech.ac.cn}
\emailAdd{duminghui@imech.ac.cn}
\emailAdd{byzhu@pmo.ac.cn}
\emailAdd{fengwf@pku.edu.cn}
\emailAdd{diaoqing23@mails.ucas.ac.cn}
\emailAdd{xupeng@imech.ac.cn}
\emailAdd{xilong.fan@whu.edu.cn}

\abstract{The measurement of the Hubble constant $H_0$ plays a central role in modern cosmology. In this work, we investigate the potential of strongly lensed gravitational-wave (SLGW) signals from massive binary black hole mergers to constrain $H_0$ using future space-based detector networks. We consider two observational scenarios: one in which the source redshift is unknown, and another in which it is independently determined through electromagnetic observations. We show that meaningful constraints on $H_0$ can still be achieved without source-redshift information, provided that the lens redshift is known. For individual SLGW events, the joint Taiji+LISA analysis improves the measurement precision of $H_0$ by approximately a factor of two compared with the Taiji-only configuration. Extending the analysis to the population level, we combine five simulated SLGW events and find that the uncertainty in $H_0$, quantified by the 95\% credible interval, reaches the $1.1\times10^{-1}$ level when the source redshift is treated as unknown, and further improves to $4.2\times10^{-2}$ when the source redshift is independently measured. Our results demonstrate that joint space-based gravitational-wave observations can substantially enhance the cosmological capability of SLGW events and provide a promising avenue for precision measurements of the Hubble constant.}

\title{Measuring the Hubble constant with strongly lensed gravitational waves from space-based detector networks}
\begin{document}
\maketitle
\flushbottom

\section{Introduction}
\label{sec:intro}

% (SNe Ia)
The Hubble constant $H_0$, which characterizes the present expansion rate of the Universe, is a cornerstone parameter in modern cosmology. In recent years, its value has been measured with ever-increasing precision using a variety of independent observational probes. However, a significant and persistent discrepancy has emerged between early- and late-Universe determinations. Measurements based on type-Ia supernovae , calibrated via the cosmic distance ladder, consistently favor a higher value of $H_0 = 73.04 \pm 1.04\,\mathrm{km\,s^{-1}\,Mpc^{-1}}$~\cite{Riess2022ApJL}, while observations of the cosmic microwave background, interpreted within the framework of the spatially flat $\Lambda$ cold dark matter ($\Lambda$CDM) model, yield a lower value of $H_0 = 67.4 \pm 0.5\,\mathrm{km\,s^{-1}\,Mpc^{-1}}$~\cite{Planck2020AA}. The discrepancy between these two benchmark measurements has reached a statistical significance of approximately $5\sigma$, and is widely referred to as the ''Hubble tension''~\cite{Riess2022ApJL, Valentino2021CQGra}. This tension may point to unaccounted systematic uncertainties in the measurements, or more intriguingly, to physics beyond the standard $\Lambda$CDM paradigm.

Ever since the first direct detection of gravitational waves (GWs) in September 2015, the LIGO-Virgo-KAGRA (LVK) collaboration has reported more than 200 GW events, with the latest GWTC--4.0 significantly expanding the sample size~\cite{Abbott2019PhRvX, Abbott2021PhRvX, Abbott2023PhRvX, Abbott2024PhRvD, LIGO2025arXiv, Jin2023ChPhC, Jin2023JCAP, Jin2024SCPMA, Song2025ApJ, Han2026EPJC, Jin2026SCPMA}. These observations have established GWs as a powerful and independent probe of the Universe. GW signals provide a direct measurement of the luminosity distance to the source, and when combined with redshift information, they can serve as standard sirens to trace the cosmic expansion history~\cite{Schutz1986Natur, Holz2005ApJ, Nissanke2013arXiv, Abbott2017Natur}. The latest LVK analysis, based on 47 events from the GWTC--3, yields a constraint of $H_0 = 68^{+8}_{-6}\,\mathrm{km\,s^{-1}\,Mpc^{-1}}$~\cite{Abbott2023ApJ}.

% (GL) 
If a GW signal from a compact binary coalescence propagates in the vicinity of massive structures, it can be gravitationally lensed in a manner analogous to electromagnetic (EM) radiation~\cite{Ohanian1974, Deguchi1986PhRvD, Wang1996PhRvL, Nakamura1998PhRvL, Takahashi2003ApJ, Li2025arXiv}. Gravitational lensing can modify the observed waveform through magnification and phase modulation, and in the strong-lensing regime, can produce multiple images with distinct arrival time delays \cite{Takahashi2003ApJ}. Such lensed GW signals provide a novel opportunity to probe cosmology, as the time delays between multiple images, together with lens modeling, enable an independent measurement of cosmological parameters~\cite{Refsdal1964MNRAS, Treu2010ARAA, Liao2017NatCo}. This approach has already been successfully applied in the EM domain, where joint analyses of strongly lensed quasars have inferred a value of $H_0 = 73.3^{+1.7}_{-1.8}\,\mathrm{km\,s^{-1}\,Mpc^{-1}}$ within the flat $\Lambda$CDM framework~\cite{Treu2016AARv, Wong2020MNRAS, Millon2020AA}.

Extensive searches for lensing signatures in GW data have been carried out using observations from the first fourth LVK observing runs (O1--O4a) (see~\cite{Hannuksela2019ApJL, Abbott2021ApJLensing, Diego2021PhRvD, Janquart2023MNRAS} and references therein), but no significant confirmed lensed GW event has been reported so far. Looking ahead, lensed GW signals are expected to be routinely observed by next-generation space-based detectors such as Taiji ~\cite{Hu2017NSRev}, TianQin~\cite{Luo2016CQGra} and LISA~\cite{Pau2017arXiv}, which will probe a population of massive binary mergers at higher redshifts~\cite{Du2026SCPMA}.

Building on these developments, strongly lensed gravitational-wave (SLGW) events have been recognized as promising cosmological probes. In particular, the time delays between multiple lensed GW signals, which are expected to be measured with high precision by future detectors, provide a direct avenue for constraining the Hubble constant ~\cite{Sereno2011MNRAS, Liao2017NatCo, Wei2017MNRAS, Li2019ApJ, Cremonese2020PDU, Hannuksela2020MNRAS, Cao2022ApJ, Hou2021MNRAS, Qi2022Univ, Ding2021MNRAS, Liu2024PhRvD}. For instance, Liao et al. \cite{Liao2017NatCo} proposed a waveform-independent framework that combines accurately measured GW time delays with EM observations of the lens system, showing that $\mathcal{O}(10)$ lensed events could measure $H_0$ to sub-percent precision within a flat $\Lambda$CDM cosmology. Despite this potential, most existing GW-based approaches to measuring $H_0$ rely on the identification of EM counterparts or host galaxies~\cite{Schutz1986Natur}. This requirement is often difficult to satisfy, particularly for high-redshift sources accessible to space-based detectors, and thus limits the applicability of such methods.

Recently, Huang et al. \cite{Huang2023JCAP} explored the use of SLGW signals observed by TianQin to measure the Hubble constant, demonstrating in particular the feasibility of approaches that do not rely on direct redshift measurements from EM observations. Space-based GW detectors, including TianQin, Taiji, and LISA, operate in partially complementary frequency bands due to differences in arm lengths and noise characteristics. TianQin achieves higher sensitivity at relatively high frequencies, whereas Taiji and LISA are optimized for lower frequencies, making them particularly well suited for detecting massive binary black holes (MBBHs) at cosmological distances. Moreover, the overlapping sensitivity of Taiji and LISA enables the formation of a coherent detector network, which can significantly improve the signal-to-noise ratio, parameter estimation accuracy, and sky localization of GW sources. These features make space-based detector networks particularly promising for high-precision $H_0$ measurements using SLGW signals. We note, however, that the analysis in Huang et al. \cite{Huang2023JCAP} is based on waveform models truncated at the innermost stable circular orbit and employs an analytic approximation to the detector response. While this treatment captures the leading-order features of the signal, further improvements incorporating more complete waveform modelling and refined detector responses may be required for robust high-precision cosmological inference.

For detected GW sources, the sky-localization precision of future space-based GW detectors is expected to reach the level of $\sim 1~\mathrm{deg}^2$ to $0.1~\mathrm{deg}^2$~\cite{Wang2019PhRvD, Liu2020PhRvD, Fan2020PhRvD, Huang2020PhRvD}, making follow-up EM observations feasible and enabling multimessenger studies of SLGW events. For an order-of-magnitude estimate of the SLGW detection rate, we adopt a lensing probability of $\sim 1\%$ for MBBH mergers observable by space-based GW detectors~\cite{Gao2022MNRAS}. Under optimistic population models, the detection rate of MBBH signals may reach $\mathcal{O}(10^2)/\mathrm{yr}$~\cite{Wang2019PhRvD, Gutierrez2025PhRvD}, implying that the total number of detectable SLGW events could be as high as $\sim5$ during a five-year mission lifetime \cite{Klein2016PhRvD, Diao2025arXiv, Yuan2026ApJ, Wang2019PhRvD, Ruan2021Resea}. The Taiji Data Challenge \cite{Du2026SCPMA} further demonstrates the feasibility of realistic signal simulations and parameter estimation pipelines for these sources.

In this work, we extend and improve upon this framework by incorporating more realistic waveform modeling and refined detector response descriptions within a multi-detector (Taiji+LISA) network configuration. We further explore the role of incomplete EM information in a systematic and self-consistent Bayesian framework. In particular, we consider two observational scenarios corresponding to different levels of available EM information. In the first scenario, the source redshift is treated as an unknown parameter and is inferred simultaneously with the source, lens, and cosmological parameters from the GW data. In the second scenario, we assume that the source redshift can be independently determined through EM observations. Such an optimistic case may be realized if the MBBH system evolves in a gas-rich environment, where gas accretion could produce detectable EM radiation associated with the GW source~\cite{Ascoli2018ApJ, Tamanini2016JCAP, Armitage2002ApJL, Milosavljevic2005ApJL}. In both scenarios, we assume that the lens redshift is known from the identification of the lensing galaxy. These two cases allow us to quantitatively assess the impact of source-redshift information on the measurement precision of the Hubble constant. As we will show, meaningful constraints on $H_0$ can be achieved even without an independent measurement of the source redshift, while the inclusion of EM redshift information significantly improves the cosmological inference.

The paper is organized as follows. Section~\ref{sec:sig} provides a brief overview of the point-mass lensing model used to describe SLGW events. In Section~\ref{sec:method}, we present the Bayesian framework for inferring the source and lens parameters, as well as the cosmological parameter $H_0$. Section~\ref{sec:sim} describes the data simulation procedure. In Section~\ref{sec:cons}, we evaluate the expected measurement precision of $H_0$ using the proposed method. Finally, Section~\ref{sec:sum} summarizes the main results and discusses possible extensions of this work. Throughout this paper, we assume a flat $\Lambda$CDM cosmology with $\Omega_m = 0.3111$, $\Omega_\Lambda = 1 - \Omega_m$, and $H_0 = 67.66~\mathrm{km\ s^{-1}\ Mpc^{-1}}$ \citep{Planck2020AA}.

\section{Strongly lensed gravitational-wave signals}
\label{sec:sig}

We simulate GW signals from MBBH mergers based on the configurations of the Taiji and LISA detectors. Throughout this work, we employ the noise-orthogonal TDI-$A_2$ and $E_2$ channels for Bayesian inference \cite{Prince2002PhRvD}. Further details of the simulation setup can be found in \cite{Yuan2026ApJ, Yuan2026ApJb}. 

The unlensed GW signal in the frequency domain is constructed as
\begin{equation}
d(f;\theta^S) = h(f;\theta^S) + n(f),
\label{eq:un-d}
\end{equation}
where $h(f;\theta^S)$ denotes the GW signal and $n(f)$ represents the detector noise. The signal can be written as
\begin{equation}
h(f;\theta^S) = h_{\tilde{A}_2,\tilde{E}_2}(f; \theta^S)
= \mathcal{T}_{A_2, E_2}(f, t_f)\, \tilde{h}(f;\theta^S),
\label{eq:tdi}
\end{equation}
where $\mathcal{T}_{A_2, E_2}(f, t_f)$ is the TDI transfer function and $\tilde{h}(f;\theta^S)$ is the intrinsic GW waveform.

In this work, we neglect black hole spins. The source parameter vector is therefore simplified to
\begin{equation}
\theta^S = (\mathcal{M}_c, q, d_\mathrm{L}, t_c, \phi_c, \iota, \lambda, \beta, \psi),
\end{equation}
where $\mathcal{M}_c$ is the chirp mass, $q$ is the mass ratio, $d_\mathrm{L}$ is the luminosity distance, $t_c$ and $\phi_c$ are the coalescence time and phase, $\iota$ is the inclination angle, $(\lambda, \beta)$ denote the ecliptic longitude and latitude, and $\psi$ is the polarization angle.

For the lens model, we adopt a point-mass approximation. The Einstein radius is defined as
\begin{equation}
\xi_0 \equiv \left(\frac{4 G M_L D_A^l D_A^{ls}}{c^2 D_A^s}\right)^{1/2},
\end{equation}
where $D_A^l$, $D_A^s$, and $D_A^{ls}$ denote the angular diameter distances from the observer to the lens, from the observer to the source, and from the lens to the source, respectively. The corresponding dimensionless lensing potential is
\begin{equation}
\psi(x) = \ln x.
\end{equation}

In the geometrical-optics limit ($w \gg 1$), the amplification factor reduces to
\begin{equation}
F(w,y) = |\mu_+|^{1/2} - i \, |\mu_-|^{1/2} e^{2 \pi i f \Delta t_d},
\end{equation}
where $w = 8 \pi G M_L (1+z_L) f / c^3$ is the dimensionless frequency. The magnifications of the two images are
\begin{equation}
\mu_\pm = \frac{1}{2} \pm \frac{y^2 + 2}{2 y \sqrt{y^2 + 4}},
\end{equation}
and the corresponding time delay is
\begin{equation}
\Delta t_d = \frac{4 G M_L (1+z_L)}{c^3} \left[ \frac{y \sqrt{y^2 + 4}}{2} + \ln \left( \frac{\sqrt{y^2 + 4} + y}{\sqrt{y^2 + 4} - y} \right) \right].
\end{equation}

The dimensionless source position is defined as
\begin{equation}
y = \frac{\eta_L c}{2 \sqrt{G M_L}} \sqrt{\frac{D_A^l}{D_A^s D_A^{ls}}},
\end{equation}
where $\eta_L$ is the transverse physical position of the source projected onto the source plane.

We adopt the standard distance--redshift relation \citep{Hogg1999astro}:
\begin{equation}
D_A(z) = \frac{1}{H_0 (1+z)} \int_0^{z} \frac{dz'}{E(z')},
\end{equation}
with
\begin{equation}
E(z) = \sqrt{\Omega_m (1+z)^3 + \Omega_\Lambda},
\end{equation}
assuming a flat $\Lambda$CDM cosmology. The luminosity distance is related to the angular diameter distance via
\begin{equation}
d_L = (1+z_s)^2 D_A^s.
\end{equation}

In the geometrical-optics limit, the lensed GW signal can be written as
\begin{equation}
h^{L}(f;\theta^S) = F(w,y)\, h(f;\theta^S).
\label{eq:hwy}
\end{equation}

For parameter inference, we reorganize the model parameters into three subsets:
\begin{align}
\theta^Z &= (\mathcal{M}_c, q, z_s, t_c, \phi_c, \iota, \lambda, \beta, \psi), \\
\theta^L &= (M_L, z_L, \eta_L), \\
\theta^H &= (H_0),
\end{align}
where $\theta^Z$, $\theta^L$, and $\theta^H$ denote the source, lens, and cosmological parameters, respectively. Accordingly, Eq.~(\ref{eq:hwy}) can be rewritten as
\begin{equation}
h^{L}(f;\theta^Z,\theta^L,\theta^H) = F(f;\theta^L)\, h(f;\theta^Z,\theta^H),
\label{eq:hL}
\end{equation}
where $F(f;\theta^L)$ is the lensing amplification factor. The corresponding observed lensed data are
\begin{equation}
d^L(f) = h^{L}(f;\theta^Z,\theta^L,\theta^H) + n(f),
\end{equation}
with $n(f)$ denoting the detector noise.

For general GW events without EM counterparts, it is typically difficult to measure the Hubble constant, as it is strongly degenerate with other source parameters, most notably the redshift \cite{Schutz1986Natur}. SLGW signals provide a mechanism to partially break this degeneracy. In particular, the lensing effect encodes information about the angular diameter distances through the lensing geometry and time-delay structure, while the GW amplitude directly probes the luminosity distance, $d_L = (1+z_s)^2 D_A^s$. Consequently, cosmological information, including $H_0$, is imprinted in the SLGW waveform.

If the lens redshift is independently measured, for example from observations of the lensing system, the degeneracy between the source redshift and cosmological parameters can be significantly alleviated, enabling a direct constraint on the Hubble constant even in the absence of an EM counterpart to the GW source.

Furthermore, when complementary EM observations are available, they can provide precise measurements of the source and lens redshifts, as well as accurate sky localization through host-galaxy identification. In this case, these quantities can be treated as known parameters, further reducing degeneracies in the parameter estimation. Accordingly, in our analysis, we consider both scenarios and, when EM information is available, we fix the redshifts of the source and lens, as well as the sky-position parameters, and infer the remaining GW and lensing parameters jointly with the Hubble constant $H_0$ from the data.

\section{Bayesian parameter estimation}
\label{sec:method}

Within a Bayesian statistical framework, we perform parameter estimation for individual SLGW events using simulated data. The posterior distribution for a single lensed event is given by
\begin{equation}
p(\theta^Z, \theta^L, H_0 \mid d^L) 
= \frac{\mathcal{L}(d^L \mid \theta^Z, \theta^L, H_0)\, p(\theta^Z, \theta^L, H_0)}
       {\mathcal{Z}^L},
\label{eq:B-lens}
\end{equation}
where $\mathcal{L}$ denotes the likelihood function, $p(\theta^Z, \theta^L, H_0)$ is the prior distribution, and $\mathcal{Z}^L$ is the Bayesian evidence,
\begin{equation}
\mathcal{Z}^L = \int \mathcal{L}(d^L \mid \theta^Z, \theta^L, H_0)\, p(\theta^Z, \theta^L, H_0)\, 
\mathrm{d}\theta^Z\, \mathrm{d}\theta^L\, \mathrm{d}H_0.
\end{equation}

We then extend the analysis to joint observations from the Taiji and LISA detectors. Denoting the lensed GW data from the two detectors as $d^L_{\mathrm{Taiji}}$ and $d^L_{\mathrm{LISA}}$, respectively, we assume that their instrumental noises are statistically independent due to their distinct orbital configurations and measurement systems. Under this assumption, the joint likelihood factorizes as
\begin{equation}
\mathcal{L}_\mathrm{joint} 
= \mathcal{L}_\mathrm{Taiji}(d^L_\mathrm{Taiji} \mid \theta^Z, \theta^L, H_0)
\times 
\mathcal{L}_\mathrm{LISA}(d^L_\mathrm{LISA} \mid \theta^Z, \theta^L, H_0).
\label{eq:taiji}
\end{equation}
The corresponding joint posterior distribution can therefore be written as
\begin{equation}
\begin{aligned}
p(\theta^Z, \theta^L, H_0 \mid d^L_\mathrm{Taiji}, d^L_\mathrm{LISA})
&\propto 
\mathcal{L}_\mathrm{Taiji}(d^L_\mathrm{Taiji} \mid \theta^Z, \theta^L, H_0) \\
&\quad \times 
\mathcal{L}_\mathrm{LISA}(d^L_\mathrm{LISA} \mid \theta^Z, \theta^L, H_0) \\
&\quad \times 
p(\theta^Z, \theta^L, H_0),
\end{aligned}
\label{eq:B-joint}
\end{equation}
where $\mathcal{L}_\mathrm{Taiji}$ and $\mathcal{L}_\mathrm{LISA}$ are the likelihood functions for the respective detectors. This multi-detector framework enables improved constraints on the Hubble constant $H_0$ by combining independent measurements.

To further enhance the constraining power, we simulate a population of jointly observed Taiji--LISA events and perform parameter estimation for each event individually. Assuming that the events are astrophysically independent and that their noise realizations are uncorrelated, the combined posterior distribution for $H_0$ from $n$ events can be expressed as
\begin{equation}
p(H_0) \propto \prod_{i=1}^{n} 
\int p(\theta^Z_i, \theta^L_i, H_0 \mid d^L_{\mathrm{Taiji},i}, d^L_{\mathrm{LISA},i}) 
\, \mathrm{d}\theta^Z_i \, \mathrm{d}\theta^L_i.
\end{equation}
This procedure allows us to quantify the statistical improvement in constraining $H_0$ achieved by combining multiple SLGW events.

\section{Data simulation}
\label{sec:sim}

Under the point-mass lens model, we simulate SLGW signals from MBBH systems according to Eq.~(\ref{eq:hL}). The unlensed GW waveforms are generated using the IMRPhenomD model~\cite{Husa2016PhRvD, Khan2016PhRvD},
which accurately describes the dominant $\ell = 2$, $|m| = 2$ modes of aligned-spin binary black hole systems. In this work, we consider only the dominant mode and neglect higher-order modes, as this approximation is sufficient to capture the main features of the SLGW signals considered here. The data simulation procedure for the Taiji mission has been described in detail in the Appendix of Refs.~\cite{Yuan2026ApJ, Yuan2026ApJb}. The simulation for the LISA mission is performed analogously, with the only modifications being the adoption of the corresponding noise sensitivity curve and detector orbit. For the LISA instrumental sensitivity and detector configuration, we
refer the reader to Ref.~\cite{Robson2019CQGra}.

During the simulation, we set the starting time of the observation segment to day 60 of the mission orbit and generate data spanning a total duration of 90 days. To ensure that the inspiral, merger, and ringdown phases are fully contained within the observation window, the coalescence time is fixed to day 149, i.e., one day before the end of the data segment. In addition, for binary systems with component masses of $10^{5}\,M_\odot$, the sampling frequency is set to $0.3\,\mathrm{Hz}$, while for systems with component masses larger than $10^{5}\,M_\odot$, the sampling frequency is reduced to $0.03\,\mathrm{Hz}$. This choice ensures sufficient waveform accuracy while improving computational efficiency.

To investigate the performance of Taiji-only observations and joint Taiji+LISA observations in constraining the Hubble constant $H_0$, we simulate a set of five multimessenger SLGW events. The simulation setup is constructed to ensure both physical consistency and observational feasibility. For a controlled comparison between the two observational scenarios, identical source, lens, and noise realizations are adopted throughout the analysis, such that any differences in the inferred constraints arise solely from the inclusion of the LISA observations rather than variations in the intrinsic event properties.

For the source population, we generate MBBH systems under a Population~III star formation scenario \cite{Vikaeus2022MNRAS,Diao2025arXiv}. The total binary mass is sampled within the range
$10^{4}\,M_\odot \leq M_{\rm tot} \leq 10^{7}\,M_\odot$, which covers the primary sensitivity band of space-based GW detectors. We further impose a mass-ratio constraint $q \geq 0.75$, motivated by theoretical predictions that
Population~III binaries preferentially form nearly
equal-mass systems. The source redshift is restricted to $z_{\rm s} \in [2,6]$, ensuring that the GW signals are detectable by both Taiji and LISA while simultaneously allowing the possibility of observable electromagnetic
counterparts from massive black holes embedded in
gas-rich environments~\cite{Ascoli2018ApJ}. This enables potential multimessenger identification with current or near-future electromagnetic facilities.

Since the coalescence time, coalescence phase, orbital inclination, polarization angle, and sky location have only a weak impact on the $H_0$ constraints considered in this work, these parameters are fixed to representative
values for all simulated events in order to reduce the computational cost. These quantities mainly correspond to extrinsic geometric degrees of freedom and exhibit only weak correlations with the cosmological parameter $H_0$. Accordingly, the analysis is conditioned on fixed values of these parameters, while the inference framework itself remains fully general and can be straightforwardly extended to the full parameter space. Specifically, we adopt
\[
(t_c, \phi_c, \lambda, \beta, \iota, \psi)
=
(149~\mathrm{day}, 4.32, 4.25, -0.15, 0.7, 0.6).
\]

For the lens population, we assume lens redshifts in the range $z_{\rm L} \in [0.2,1]$, corresponding to typical galaxy-scale lenses at intermediate cosmological distances. The lens mass is sampled uniformly in logarithmic space between $10^{9}\,M_\odot$ and $10^{12}\,M_\odot$, covering the characteristic mass range of galaxy and massive-galaxy lenses responsible for strong gravitational lensing. Finally, to ensure that each simulated system produces at
least two lensed images, we uniformly sample the
dimensionless lensing parameter $\eta_{\rm L}$ within the interval $[1,5]$. This selection excludes single-image configurations and guarantees that the corresponding time delays and magnification factors fall within a regime that can be resolved and analyzed in joint Taiji--LISA
observations.

\section{Measurement of the Hubble Constant}
\label{sec:cons}

Among the 5 simulated events, we select one representative example to demonstrate the constraints on
the Hubble constant $H_0$. The source parameters of this event are given by
$\mathcal{M}_{c} = 3513848.7\,M_\odot$,
$q = 0.77$ and $z_s = 4.8$.
The corresponding lens parameters are
$\eta_{\rm L} = 3.4$,
$M_{\rm L} = 1.0 \times 10^{11}\,M_\odot$,
and
$z_{\rm L} = 0.6$.
% For this simulated event, we adopt $H_0 = 67.66\,\mathrm{km\,s^{-1}\,Mpc^{-1}}$ as the injected cosmological value.

\begin{figure}[htbp]
\centering
\includegraphics[width=15cm]{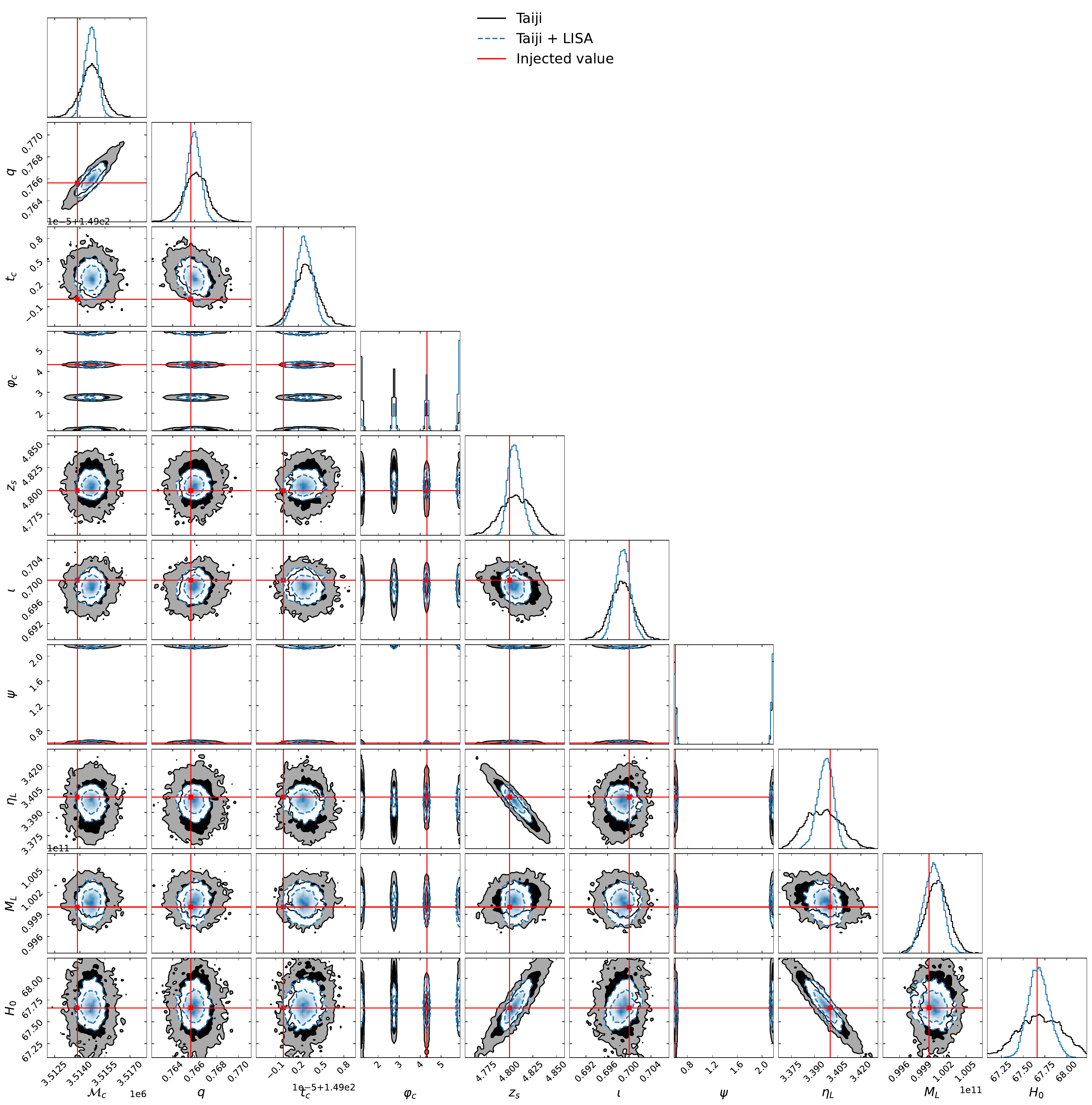}
\caption{Posterior distributions of the cosmological and lensing parameters for the case with free source redshift. The black contours correspond to the Taiji-only results, while the blue contours represent the joint Taiji--LISA analysis. The red lines indicate the injected values of the parameters.}
\label{case11_zs}
\end{figure}

\begin{figure}[htbp]
\centering
\includegraphics[width=15cm]{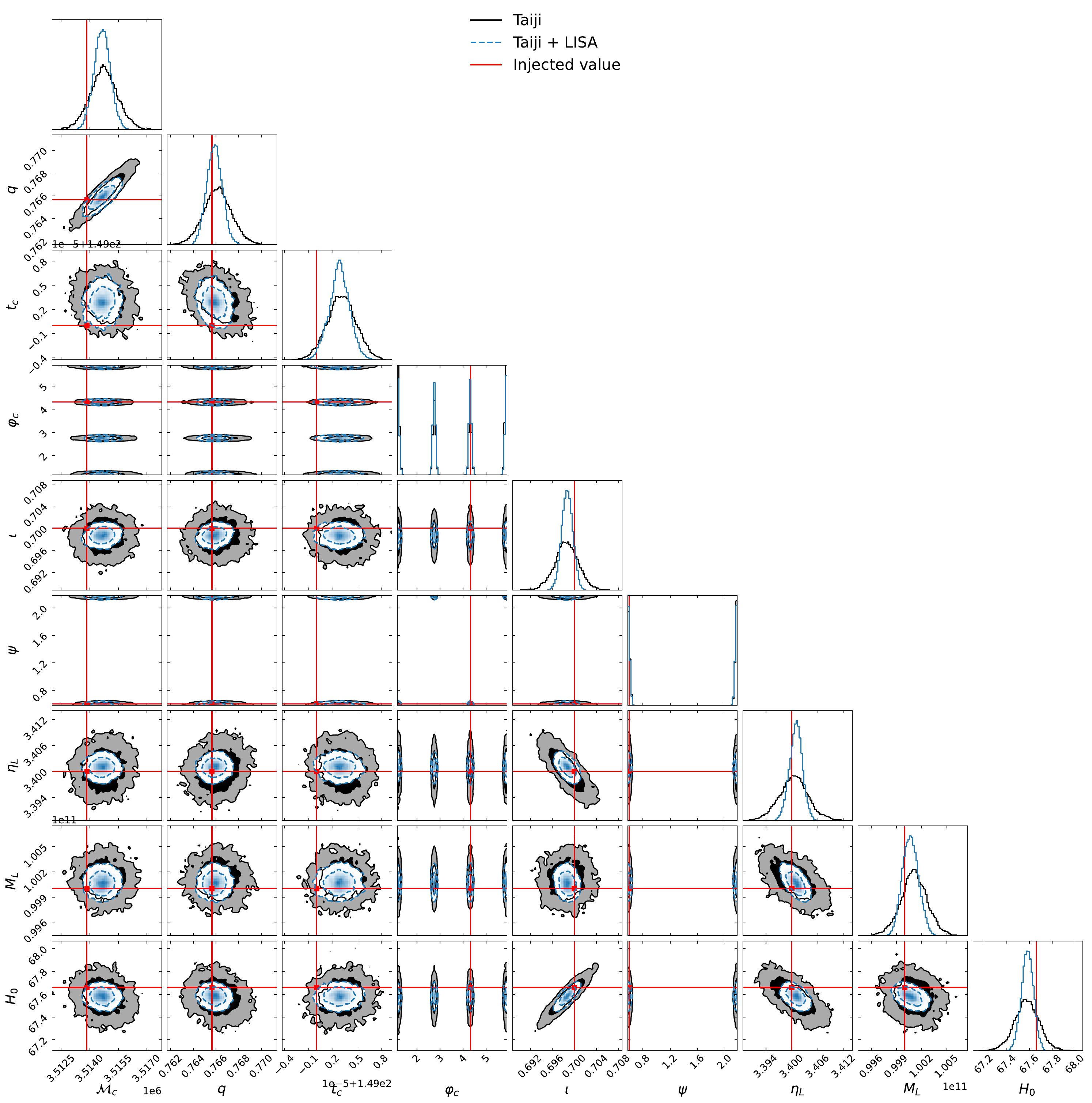}
\caption{Same as Fig.~\ref{case11_zs}, but for the case with fixed source redshift.}
\label{case11_withoutzs}
\end{figure}

Using the Bayesian posterior defined in Eq.~(\ref{eq:taiji}) for the single-detector analysis and Eq.~(\ref{eq:B-joint}) for the joint Taiji--LISA analysis, we perform parameter estimation for this event under two
different observational assumptions regarding the source
redshift. In the first scenario, the source redshift is treated as an unknown parameter and is inferred simultaneously with the source parameters, lens parameters, and the Hubble constant $H_0$. The resulting posterior distributions are shown in figure~\ref{case11_zs}. In the second scenario, we assume that the source redshift can be independently determined from EM observations and therefore fix it to the injected value. Under this assumption, we jointly infer the remaining source parameters, lens parameters, and the Hubble constant $H_0$. The corresponding posterior distributions are presented in figure~\ref{case11_withoutzs}.

For the GW source parameters, the joint Taiji–LISA analysis yields consistently improved measurement precision across nearly all dimensions, irrespective of whether the source redshift is fixed or treated as a free parameter. Compared with the Taiji-only case, the one-dimensional posterior distributions of the chirp mass $\mathcal{M}_c$ and the mass ratio $q$ become markedly narrower, with their 95\% credible intervals significantly reduced. This improvement is primarily attributed to the enhanced network signal-to-noise ratio and the partial breaking of parameter degeneracies enabled by the multi-detector configuration.
Strong correlations present in several two-dimensional posterior distributions are also substantially weakened. These results indicate that the complementary antenna response patterns and frequency-dependent sensitivities of Taiji and LISA effectively reduce parameter degeneracies and improve the overall fidelity of source parameter inference.

In contrast, the intrinsic degeneracy between the coalescence phase $\phi_c$ and the polarization angle $\psi$ is not fully resolved even in the joint Taiji–LISA analysis. For all 5 events considered, the posterior distributions in the $(\phi_c, \psi)$ plane remain multimodal. This behavior originates from an approximate degeneracy along directions of constant $\phi_c + \psi$ and $\phi_c - \psi$, leading to mirror solutions in the joint parameter space. Consequently, although the inclusion of LISA improves global parameter constraints, it is insufficient to break this specific degeneracy. This limitation suggests that incorporating higher-order modes in the waveform model would be necessary to fully resolve the ambiguity, as demonstrated in previous studies (e.g. \cite{Marsat2021PhRvD, Yuan2026ApJb}).

The improvement is even more pronounced for the lensing parameters. The posterior distributions of the lens mass $M_L$ and the dimensionless lensing parameter $\eta_L$ obtained from the joint Taiji–LISA analysis are significantly tighter than those from Taiji alone, with uncertainties reduced by nearly an order of magnitude. Moreover, the correlations between $M_L$ and the intrinsic source parameters are substantially suppressed, demonstrating that the joint observation strategy is considerably more effective in disentangling intrinsic waveform properties from lensing-induced magnification effects.

Among the cosmological parameters, the constraint on the Hubble constant $H_0$ shows the most substantial improvement when LISA observations are incorporated. When the source redshift is treated as an unknown parameter, the Taiji-only analysis yields a constraint of $H_0 = 67.69^{+0.45}_{-0.47}$, while the joint Taiji--LISA analysis improves this to $H_0 = 67.68^{+0.25}_{-0.21}$, where all quoted uncertainties correspond to the 95\% confidence interval. The corresponding posterior distributions are shown in Fig.~\ref{case11_zs}.
In the case where the source redshift can be independently determined from EM observations, the constraints are further improved to $H_0 = 67.58^{+0.22}_{-0.22}$ for Taiji alone and $H_0 = 67.58^{+0.10}_{-0.10}$ for the joint Taiji--LISA configuration, as illustrated in Fig.~\ref{case11_withoutzs}. Compared with the scenario in which the source redshift is treated as a free parameter, the availability of an independent EM redshift measurement further improves the precision of the $H_0$ determination by approximately a factor of 2 for both the Taiji-only and joint Taiji--LISA analyses. This improvement arises from breaking the degeneracy between the source redshift and the cosmological parameters in the gravitational-wave inference. In both scenarios, the inclusion of LISA significantly enhances the constraining power of the detector network, leading to a substantial reduction in the width of the credible interval. In particular, the joint Taiji--LISA analysis improves the precision of the $H_0$ measurement by approximately a factor of 2 compared with the Taiji-only case. These results demonstrate the potential of SLGW events observed by space-based detector networks to provide an independent measurement of the Hubble constant in the corresponding redshift range.

In addition to the improved one-dimensional constraint, the joint Taiji--LISA analysis also significantly suppresses the correlations between the Hubble constant $H_0$ and the lensing parameters, particularly the lens mass $M_{\mathrm{L}}$ and the dimensionless lensing parameter $\eta_{\mathrm{L}}$. The reduction of these parameter degeneracies indicates that the multidetector configuration is more effective in disentangling the cosmological information encoded in $H_0$ from lensing-induced systematics that are inherent in single-detector observations.

\begin{figure}[htbp]
\centering
\includegraphics[width=7cm]{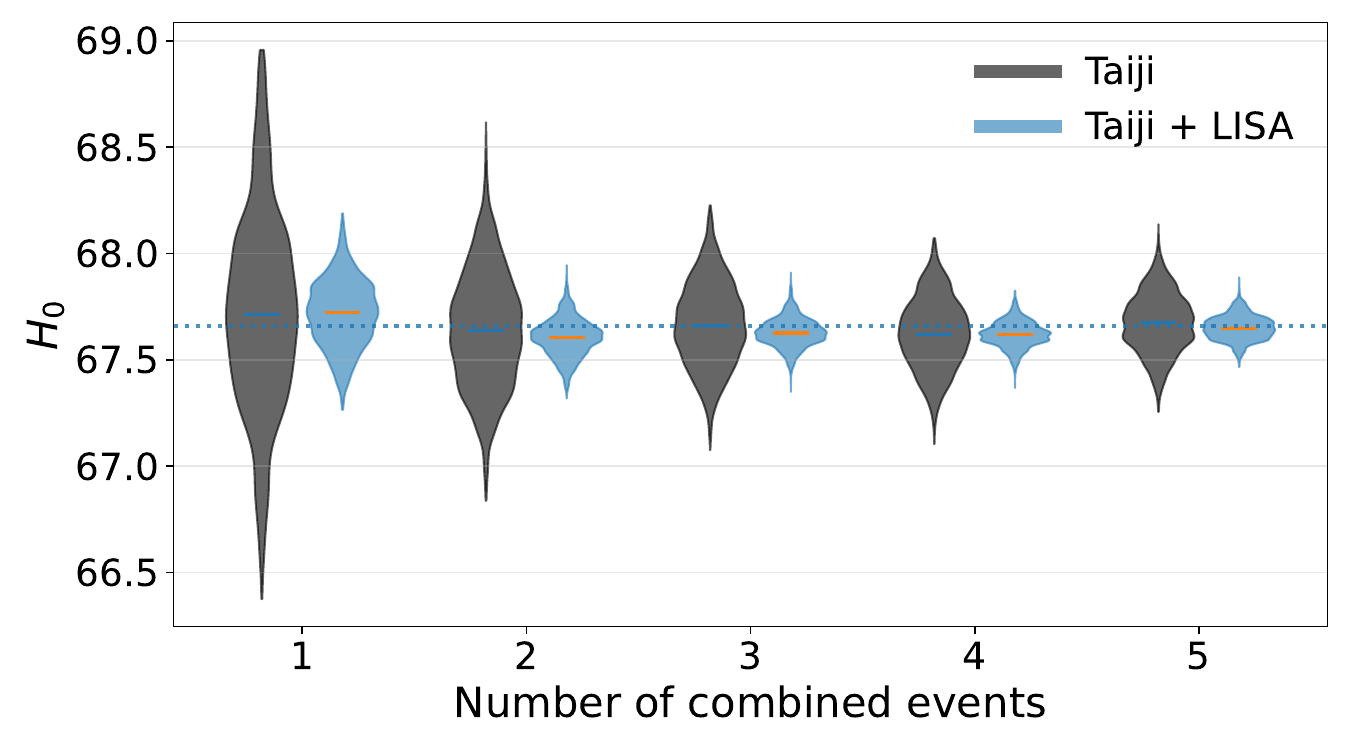}
\includegraphics[width=7cm]{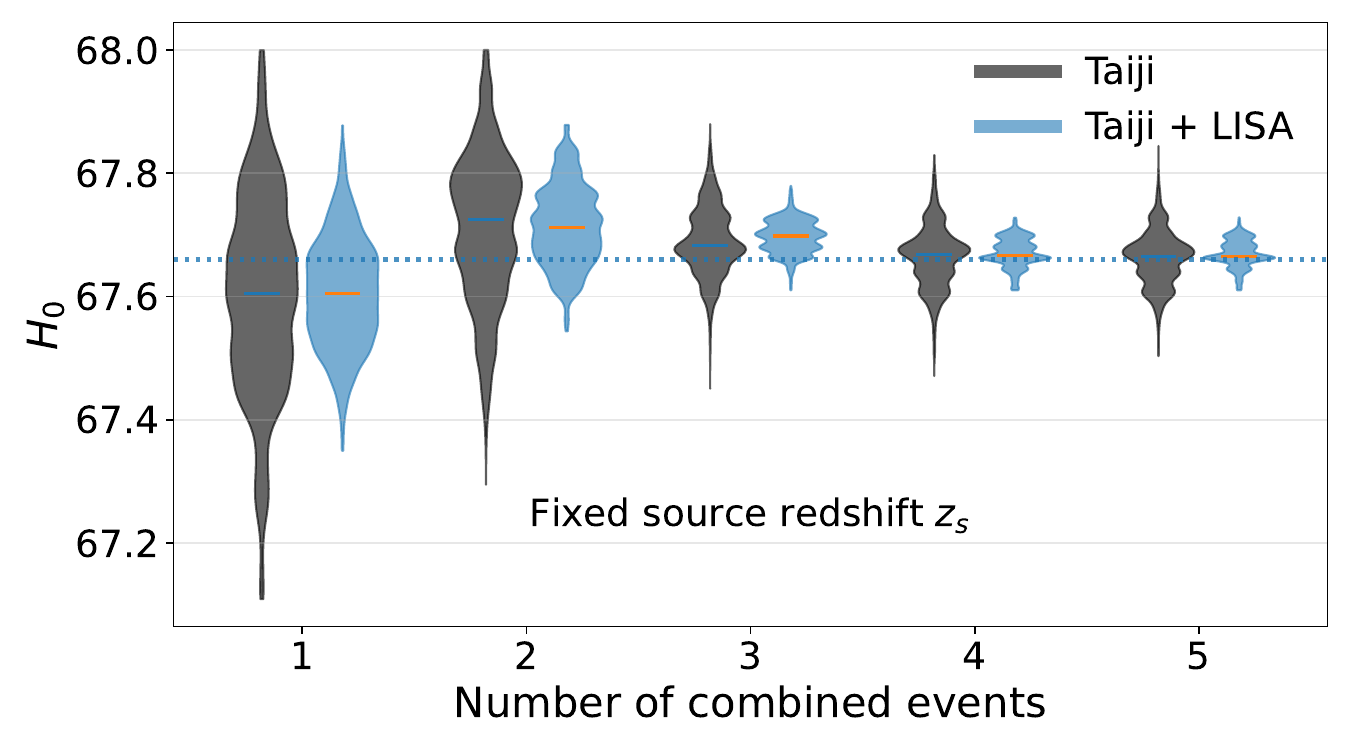}
\caption{Population-level constraints on the Hubble constant $H_0$ obtained by progressively combining five SLGW events. Left panel: results assuming the case where the source redshift is treated as a free parameter. Right panel: results assuming independently determined source redshifts. Blue and orange violins correspond to Taiji-only and Taiji+LISA observations, respectively. The horizontal dotted line indicates the fiducial value of $H_0$. In all cases, the population posteriors remain consistent with the fiducial value, showing no evidence of bias in the recovery of $H_0$ and demonstrating the robustness of the inference.}
\label{H0_pop}
\end{figure}

Figure~\ref{H0_pop} displays the population posterior distributions of the Hubble constant $H_0$ inferred from the sequential combination of up to five SLGW. The left panel corresponds to the case in which the source redshift is treated as an unknown parameter, while the right panel shows the results obtained when the source redshift is independently determined and fixed from EM observations. For each number of combined events, the violin plots represent the reconstructed population posterior of $H_0$, obtained by multiplying the posteriors from individual events.

For both cases, namely with and without an independent EM determination of the source redshift, the population posteriors remain fully consistent with the fiducial value of the Hubble constant adopted in the simulation, indicating no statistically significant bias in the recovery of $H_0$. Furthermore, as additional SLGW events are sequentially incorporated into the analysis, the posterior distributions become progressively narrower, demonstrating the expected cumulative improvement in the constraining power of the population analysis. The reduction in the posterior width is more pronounced in the scenario where the source redshift is fixed by EM observations, highlighting the importance of redshift information in improving the precision of cosmological parameter estimation.

After combining 5 SLGW events, we find that the $95\%$ credible uncertainty on the Hubble constant is significantly reduced through the population analysis. In the case where the source redshift is treated as an unknown parameter, the uncertainty on $H_0$ is constrained to $2.7\times10^{-1}$ for the Taiji-only configuration and further reduced to $1.1\times10^{-1}$ for the joint Taiji--LISA analysis. When the source redshift is independently determined from EM observations, the corresponding uncertainties improve to $8.5\times10^{-2}$ and $4.2\times10^{-2}$ for the Taiji-only and Taiji--LISA configurations, respectively.

Comparing the two observational scenarios, the case with independently measured source redshifts consistently yields tighter constraints on $H_0$ than the case where the redshift is left free, owing to the removal of the degeneracy between redshift and cosmological parameters in the GW inference. In addition, the joint Taiji--LISA configuration systematically provides more stringent population-level constraints than the Taiji-only case. This improvement originates from the enhanced parameter estimation accuracy achieved for individual events when LISA observations are incorporated, which coherently propagates through the population combination.

Overall, these results demonstrate the clear advantage of joint space-based GW observations for population studies of strongly lensed events and highlight their strong potential for precision measurements of the Hubble constant.

\section{Summary and discussion}
\label{sec:sum}

In this work, we analyze the GW signals from MBBHs expected in the Taiji and LISA frequency bands under the point-mass lens model, focusing on the impact of strong lensing that produces two distinct images in the observations. To this end, we construct 5 simulated events based on the Population III stellar formation model and generate 90 days of synthetic GW data for both Taiji and LISA. The source redshifts are distributed in the range $z_s \in [2, 6]$, while the lens redshifts lie within $z_L \in [0.2, 1]$. We perform parameter estimation using Taiji-only data as well as joint Taiji+LISA observations to measure the Hubble constant $H_0$. Furthermore, we assess the combined constraining power of the 5 simulated events, quantifying the achievable precision in the measurement of $H_0$.

For individual lensed events, as shown in Figures \ref{case11_zs} and \ref{case11_withoutzs}, the joint Taiji+LISA analysis consistently achieves higher precision in the measurement of the Hubble constant $H_0$ compared with the Taiji-only configuration. In particular, the one-dimensional posterior distributions of the source chirp mass $\mathcal{M}_c$, mass ratio $q$, and coalescence time $t_c$ are significantly narrower, and degeneracies visible in the two-dimensional posteriors are substantially mitigated. Similarly, the lens mass $M_{\mathrm{L}}$ and the dimensionless lensing parameter $\eta_{\mathrm{L}}$ are more tightly constrained, with uncertainties reduced by nearly an order of magnitude in the joint analysis. For the Hubble constant $H_0$, under the case where the source redshift is treated as an unknown parameter, the Taiji-only analysis yields a constraint of $67.69^{+0.45}_{-0.47}$, while the joint Taiji+LISA analysis gives $H_0 = 67.68^{+0.25}_{-0.21}$. In both cases, the inferred values of $H_0$ remain consistent with the fiducial value within the quoted uncertainties, indicating no evidence for bias in the recovery of $H_0$ at the level of precision achieved. When the source redshift is independently determined from EM observations, the corresponding constraints are further improved to $H_0 = 67.58^{+0.22}_{-0.22}$ for Taiji alone and $H_0 = 67.58^{+0.10}_{-0.10}$ for the joint Taiji+LISA analysis. All quoted uncertainties correspond to the 95\% confidence intervals. In both scenarios, the inclusion of LISA reduces the credible-interval width and suppresses correlations between $H_0$ and the lensing parameters, demonstrating the enhanced ability of the multidetector configuration to disentangle cosmological information from lensing-induced systematics.

Extending the analysis to the population level, we sequentially combine up to 5 simulated SLGW events to quantify the cumulative improvement in the measurement of the Hubble constant $H_0$. The resulting population posteriors for the two adopted redshift scenarios are shown in Fig.~\ref{H0_pop}. When the source redshift is treated as a free parameter, the Taiji-only analysis constrains $H_0$ with a $95\%$ credible uncertainty of $2.7\times10^{-1}$, while the joint Taiji+LISA analysis improves the constraint by approximately a factor of two. If the source redshift is independently determined through EM observations, the uncertainty is further reduced to the $4.2\times10^{-2}$ level for the combined Taiji+LISA configuration, again representing an improvement of roughly a factor of two relative to the Taiji-only case. In all scenarios considered, the recovered population posteriors remain consistent with the fiducial value of $H_0$, indicating no significant systematic bias in the inferred cosmological parameter across the simulated event set. These results demonstrate that joint space-based GW detector networks can substantially enhance the cosmological potential of strongly lensed GW observations, particularly when complementary EM information is available. A direct comparison between the two redshift assumptions shows that independently measured source redshifts lead to consistently tighter constraints on $H_0$, reflecting the removal of the degeneracy between redshift and luminosity distance in the inference. Moreover, for both assumptions, the Taiji+LISA configuration systematically outperforms the Taiji-only case, highlighting the crucial role of multidetector synergy in improving individual-event parameter estimation and coherently propagating these gains to the population level. Overall, our results demonstrate the strong potential of multi-space-based GW detector networks for precision measurements of the Hubble constant using SLGW events.

Finally, we emphasize that our analysis is conducted solely within the point-mass lens model, and that the simulated data used in this study are based on idealized assumptions. Specifically, we consider only GW signals from MBBHs and assume the presence of a single lens along the line of sight, neglecting the potential effects of multiple lenses or overlapping sources. In addition, we do not take into account the impact of weak gravitational lensing, which may introduce additional stochastic fluctuations in the observed luminosity distance and thereby affect parameter inference in realistic analyses. In actual Taiji and LISA observations, additional GW signals from other populations, such as double white dwarfs, would also be present and may contribute to the overall data stream. Future work will aim to extend our analysis to lensing signals embedded in more realistic observational data, including the effects of weak lensing, multiple lenses, and overlapping sources, thereby providing a closer representation of actual space-based GW measurements.

% \appendix
% \section{Some title}
% Please always give a title also for appendices.

\acknowledgments
We thank Shun-Jia Huang and Huan Zhou for helpful
discussions.
This work is supported by National Key Research and Development Program of China, No. 2025YFE0217300. W.-F.F. acknowledges support from the China Postdoctoral Science Foundation (Grant No. 2025M783222).

% Bibliography

%% [A] Recommended: using JHEP.bst file
\bibliographystyle{JHEP}
\bibliography{biblio.bib}

@ARTICLE{Takahashi2003ApJ,
       author = {{Takahashi}, Ryuichi and {Nakamura}, Takashi},
        title = "{Wave Effects in the Gravitational Lensing of Gravitational Waves from Chirping Binaries}",
      journal = {\apj},
         year = 2003,
        month = oct,
       volume = {595},
       number = {2},
        pages = {1039-1051},
          doi = {10.1086/377430},
archivePrefix = {arXiv},
       eprint = {astro-ph/0305055},
 primaryClass = {astro-ph},
       adsurl = {https://ui.adsabs.harvard.edu/abs/2003ApJ...595.1039T}
}

@ARTICLE{Prince2002PhRvD,
       author = {{Prince}, Thomas A. and {Tinto}, Massimo and {Larson}, Shane L. and {Armstrong}, J.~W.},
        title = "{LISA optimal sensitivity}",
      journal = {\prd},
         year = 2002,
        month = dec,
       volume = {66},
       number = {12},
          eid = {122002},
        pages = {122002},
          doi = {10.1103/PhysRevD.66.122002},
archivePrefix = {arXiv},
       eprint = {gr-qc/0209039},
 primaryClass = {gr-qc},
       adsurl = {https://ui.adsabs.harvard.edu/abs/2002PhRvD..66l2002P}
}

@ARTICLE{Yuan2026ApJ,
       author = {{Yuan}, Yong and {Du}, Minghui and {Lin}, Xin-yi and {Zhou}, Huan and {Xu}, Peng and {Fan}, Xilong},
        title = "{Bayesian Analysis of Wave-optics Gravitationally Lensed Massive Black Hole Binaries with a Space-based Gravitational-wave Detector}",
      journal = {\apj},
         year = 2026,
        month = jan,
       volume = {997},
       number = {1},
          eid = {11},
        pages = {11},
          doi = {10.3847/1538-4357/ae29ad},
archivePrefix = {arXiv},
       eprint = {2509.01888},
 primaryClass = {astro-ph.HE},
       adsurl = {https://ui.adsabs.harvard.edu/abs/2026ApJ...997...11Y}
}

@ARTICLE{Hogg1999astro,
       author = {{Hogg}, David W.},
        title = "{Distance measures in cosmology}",
      journal = {arXiv e-prints},
         year = 1999,
        month = may,
          eid = {astro-ph/9905116},
        pages = {astro-ph/9905116},
          doi = {10.48550/arXiv.astro-ph/9905116},
archivePrefix = {arXiv},
       eprint = {astro-ph/9905116},
 primaryClass = {astro-ph},
       adsurl = {https://ui.adsabs.harvard.edu/abs/1999astro.ph..5116H}
}

@ARTICLE{Schutz1986Natur,
       author = {{Schutz}, B.~F.},
        title = "{Determining the Hubble constant from gravitational wave observations}",
      journal = {\nat},
         year = 1986,
        month = sep,
       volume = {323},
       number = {6086},
        pages = {310-311},
          doi = {10.1038/323310a0},
       adsurl = {https://ui.adsabs.harvard.edu/abs/1986Natur.323..310S}
}

@ARTICLE{Husa2016PhRvD,
       author = {{Husa}, Sascha and {Khan}, Sebastian and {Hannam}, Mark and {P{\"u}rrer}, Michael and {Ohme}, Frank and {Forteza}, Xisco Jim{\'e}nez and {Boh{\'e}}, Alejandro},
        title = "{Frequency-domain gravitational waves from nonprecessing black-hole binaries. I. New numerical waveforms and anatomy of the signal}",
      journal = {\prd},
         year = 2016,
        month = feb,
       volume = {93},
       number = {4},
          eid = {044006},
        pages = {044006},
          doi = {10.1103/PhysRevD.93.044006},
archivePrefix = {arXiv},
       eprint = {1508.07250},
 primaryClass = {gr-qc},
       adsurl = {https://ui.adsabs.harvard.edu/abs/2016PhRvD..93d4006H}
}

@ARTICLE{Holz2005ApJ,
       author = {{Holz}, Daniel E. and {Hughes}, Scott A.},
        title = "{Using Gravitational-Wave Standard Sirens}",
      journal = {\apj},
         year = 2005,
        month = aug,
       volume = {629},
       number = {1},
        pages = {15-22},
          doi = {10.1086/431341},
archivePrefix = {arXiv},
       eprint = {astro-ph/0504616},
 primaryClass = {astro-ph},
       adsurl = {https://ui.adsabs.harvard.edu/abs/2005ApJ...629...15H}
}

@ARTICLE{Nissanke2013arXiv,
       author = {{Nissanke}, Samaya and {Holz}, Daniel E. and {Dalal}, Neal and {Hughes}, Scott A. and {Sievers}, Jonathan L. and {Hirata}, Christopher M.},
        title = "{Determining the Hubble constant from gravitational wave observations of merging compact binaries}",
      journal = {arXiv e-prints},
         year = 2013,
        month = jul,
          eid = {arXiv:1307.2638},
        pages = {arXiv:1307.2638},
          doi = {10.48550/arXiv.1307.2638},
archivePrefix = {arXiv},
       eprint = {1307.2638},
 primaryClass = {astro-ph.CO},
       adsurl = {https://ui.adsabs.harvard.edu/abs/2013arXiv1307.2638N}
}

@ARTICLE{Abbott2017Natur,
       author = {{Abbott}, B.~P. and {Abbott}, R. and {Abbott}, T.~D. and {Acernese}, F. and {Ackley}, K. and {Adams}, C. and {Adams}, T. and {Addesso}, P. and {Adhikari}, R.~X. and {Adya}, V.~B. and {Affeldt}, C. and {Afrough}, M. and {Agarwal}, B. and {Agathos}, M. and {Agatsuma}, K. and {Aggarwal}, N. and {Aguiar}, O.~D. and {Aiello}, L. and {Ain}, A. and {Ajith}, P. and {Allen}, B. and {Allen}, G. and {Allocca}, A. and {Altin}, P.~A. and {Amato}, A. and {Ananyeva}, A. and {Anderson}, S.~B. and {Anderson}, W.~G. and {Angelova}, S.~V. and {Antier}, S. and {Appert}, S. and {Arai}, K. and {Araya}, M.~C. and {Areeda}, J.~S. and {Arnaud}, N. and {Arun}, K.~G. and {Ascenzi}, S. and {Ashton}, G. and {Ast}, M. and {Aston}, S.~M. and {Astone}, P. and {Atallah}, D.~V. and {Aufmuth}, P. and {Aulbert}, C. and {Aultoneal}, K. and {Austin}, C. and {Avila-Alvarez}, A. and {Babak}, S. and {Bacon}, P. and {Bader}, M.~K.~M. and {Bae}, S. and {Baker}, P.~T. and {Baldaccini}, F. and {Ballardin}, G. and {Ballmer}, S.~W. and {Banagiri}, S. and {Barayoga}, J.~C. and {Barclay}, S.~E. and {Barish}, B.~C. and {Barker}, D. and {Barkett}, K. and {Barone}, F. and {Barr}, B. and {Barsotti}, L. and {Barsuglia}, M. and {Barta}, D. and {Bartlett}, J. and {Bartos}, I. and {Bassiri}, R. and {Basti}, A. and {Batch}, J.~C. and {Bawaj}, M. and {Bayley}, J.~C. and {Bazzan}, M. and {B{\'e}csy}, B. and {Beer}, C. and {Bejger}, M. and {Belahcene}, I. and {Bell}, A.~S. and {Berger}, B.~K. and {Bergmann}, G. and {Bero}, J.~J. and {Berry}, C.~P.~L. and {Bersanetti}, D. and {Bertolini}, A. and {Betzwieser}, J. and {Bhagwat}, S. and {Bhandare}, R. and {Bilenko}, I.~A. and {Billingsley}, G. and {Billman}, C.~R. and {Birch}, J. and {Birney}, R. and {Birnholtz}, O. and {Biscans}, S. and {Biscoveanu}, S. and {Bisht}, A. and {Bitossi}, M. and {Biwer}, C. and {Bizouard}, M.~A. and {Blackburn}, J.~K. and {Blackman}, J. and {Blair}, C.~D. and {Blair}, D.~G. and {Blair}, R.~M. and {Bloemen}, S. and {Bock}, O. and {Bode}, N. and {Boer}, M. and {Bogaert}, G. and {Bohe}, A. and {Bondu}, F. and {Bonilla}, E. and {Bonnand}, R. and {Boom}, B.~A. and {Bork}, R. and {Boschi}, V. and {Bose}, S. and {Bossie}, K. and {Bouffanais}, Y. and {Bozzi}, A. and {Bradaschia}, C. and {Brady}, P.~R. and {Branchesi}, M. and {Brau}, J.~E. and {Briant}, T. and {Brillet}, A. and {Brinkmann}, M. and {Brisson}, V. and {Brockill}, P. and {Broida}, J.~E. and {Brooks}, A.~F. and {Brown}, D.~A. and {Brown}, D.~D. and {Brunett}, S. and {Buchanan}, C.~C. and {Buikema}, A. and {Bulik}, T. and {Bulten}, H.~J. and {Buonanno}, A. and {Buskulic}, D. and {Buy}, C. and {Byer}, R.~L. and {Cabero}, M. and {Cadonati}, L. and {Cagnoli}, G. and {Cahillane}, C. and {Bustillo}, J. Calder{\'o}n and {Callister}, T.~A. and {Calloni}, E. and {Camp}, J.~B. and {Canepa}, M. and {Canizares}, P. and {Cannon}, K.~C. and {Cao}, H. and {Cao}, J. and {Capano}, C.~D. and {Capocasa}, E. and {Carbognani}, F. and {Caride}, S. and {Carney}, M.~F. and {Diaz}, J. Casanueva and {Casentini}, C. and {Caudill}, S. and {Cavagli{\`a}}, M. and {Cavalier}, F. and {Cavalieri}, R. and {Cella}, G. and {Cepeda}, C.~B. and {Cerd{\'a}-Dur{\'a}n}, P. and {Cerretani}, G. and {Cesarini}, E. and {Chamberlin}, S.~J. and {Chan}, M. and {Chao}, S. and {Charlton}, P. and {Chase}, E. and {Chassande-Mottin}, E. and {Chatterjee}, D. and {Chatziioannou}, K. and {Cheeseboro}, B.~D. and {Chen}, H.~Y. and {Chen}, X. and {Chen}, Y. and {Cheng}, H.-P. and {Chia}, H. and {Chincarini}, A. and {Chiummo}, A. and {Chmiel}, T. and {Cho}, H.~S. and {Cho}, M. and {Chow}, J.~H. and {Christensen}, N. and {Chu}, Q. and {Chua}, A.~J.~K. and {Chua}, S. and {Chung}, A.~K.~W. and {Chung}, S. and {Ciani}, G. and {Ciolfi}, R.},
        title = "{A gravitational-wave standard siren measurement of the Hubble constant}",
      journal = {\nat},
         year = 2017,
        month = nov,
       volume = {551},
       number = {7678},
        pages = {85-88},
          doi = {10.1038/nature24471},
archivePrefix = {arXiv},
       eprint = {1710.05835},
 primaryClass = {astro-ph.CO},
       adsurl = {https://ui.adsabs.harvard.edu/abs/2017Natur.551...85A}
}

@ARTICLE{Khan2016PhRvD,
       author = {{Khan}, Sebastian and {Husa}, Sascha and {Hannam}, Mark and {Ohme}, Frank and {P{\"u}rrer}, Michael and {Forteza}, Xisco Jim{\'e}nez and {Boh{\'e}}, Alejandro},
        title = "{Frequency-domain gravitational waves from nonprecessing black-hole binaries. II. A phenomenological model for the advanced detector era}",
      journal = {\prd},
         year = 2016,
        month = feb,
       volume = {93},
       number = {4},
          eid = {044007},
        pages = {044007},
          doi = {10.1103/PhysRevD.93.044007},
archivePrefix = {arXiv},
       eprint = {1508.07253},
 primaryClass = {gr-qc},
       adsurl = {https://ui.adsabs.harvard.edu/abs/2016PhRvD..93d4007K}
}

@ARTICLE{Robson2019CQGra,
       author = {{Robson}, Travis and {Cornish}, Neil J. and {Liu}, Chang},
        title = "{The construction and use of LISA sensitivity curves}",
      journal = {Classical and Quantum Gravity},
         year = 2019,
        month = may,
       volume = {36},
       number = {10},
          eid = {105011},
        pages = {105011},
          doi = {10.1088/1361-6382/ab1101},
archivePrefix = {arXiv},
       eprint = {1803.01944},
 primaryClass = {astro-ph.HE},
       adsurl = {https://ui.adsabs.harvard.edu/abs/2019CQGra..36j5011R}
}

@ARTICLE{Riess2022ApJL,
       author = {{Riess}, Adam G. and {Yuan}, Wenlong and {Macri}, Lucas M. and {Scolnic}, Dan and {Brout}, Dillon and {Casertano}, Stefano and {Jones}, David O. and {Murakami}, Yukei and {Anand}, Gagandeep S. and {Breuval}, Louise and {Brink}, Thomas G. and {Filippenko}, Alexei V. and {Hoffmann}, Samantha and {Jha}, Saurabh W. and {D'arcy Kenworthy}, W. and {Mackenty}, John and {Stahl}, Benjamin E. and {Zheng}, WeiKang},
        title = "{A Comprehensive Measurement of the Local Value of the Hubble Constant with 1 km s$^{-1}$ Mpc$^{-1}$ Uncertainty from the Hubble Space Telescope and the SH0ES Team}",
      journal = {\apjl},
         year = 2022,
        month = jul,
       volume = {934},
       number = {1},
          eid = {L7},
        pages = {L7},
          doi = {10.3847/2041-8213/ac5c5b},
archivePrefix = {arXiv},
       eprint = {2112.04510},
 primaryClass = {astro-ph.CO},
       adsurl = {https://ui.adsabs.harvard.edu/abs/2022ApJ...934L...7R}
}

@ARTICLE{Planck2020AA,
       author = {{Planck Collaboration} and {Aghanim}, N. and {Akrami}, Y. and {Ashdown}, M. and {Aumont}, J. and {Baccigalupi}, C. and {Ballardini}, M. and {Banday}, A.~J. and {Barreiro}, R.~B. and {Bartolo}, N. and {Basak}, S. and {Battye}, R. and {Benabed}, K. and {Bernard}, J.-P. and {Bersanelli}, M. and {Bielewicz}, P. and {Bock}, J.~J. and {Bond}, J.~R. and {Borrill}, J. and {Bouchet}, F.~R. and {Boulanger}, F. and {Bucher}, M. and {Burigana}, C. and {Butler}, R.~C. and {Calabrese}, E. and {Cardoso}, J.-F. and {Carron}, J. and {Challinor}, A. and {Chiang}, H.~C. and {Chluba}, J. and {Colombo}, L.~P.~L. and {Combet}, C. and {Contreras}, D. and {Crill}, B.~P. and {Cuttaia}, F. and {de Bernardis}, P. and {de Zotti}, G. and {Delabrouille}, J. and {Delouis}, J.-M. and {Di Valentino}, E. and {Diego}, J.~M. and {Dor{\'e}}, O. and {Douspis}, M. and {Ducout}, A. and {Dupac}, X. and {Dusini}, S. and {Efstathiou}, G. and {Elsner}, F. and {En{\ss}lin}, T.~A. and {Eriksen}, H.~K. and {Fantaye}, Y. and {Farhang}, M. and {Fergusson}, J. and {Fernandez-Cobos}, R. and {Finelli}, F. and {Forastieri}, F. and {Frailis}, M. and {Fraisse}, A.~A. and {Franceschi}, E. and {Frolov}, A. and {Galeotta}, S. and {Galli}, S. and {Ganga}, K. and {G{\'e}nova-Santos}, R.~T. and {Gerbino}, M. and {Ghosh}, T. and {Gonz{\'a}lez-Nuevo}, J. and {G{\'o}rski}, K.~M. and {Gratton}, S. and {Gruppuso}, A. and {Gudmundsson}, J.~E. and {Hamann}, J. and {Handley}, W. and {Hansen}, F.~K. and {Herranz}, D. and {Hildebrandt}, S.~R. and {Hivon}, E. and {Huang}, Z. and {Jaffe}, A.~H. and {Jones}, W.~C. and {Karakci}, A. and {Keih{\"a}nen}, E. and {Keskitalo}, R. and {Kiiveri}, K. and {Kim}, J. and {Kisner}, T.~S. and {Knox}, L. and {Krachmalnicoff}, N. and {Kunz}, M. and {Kurki-Suonio}, H. and {Lagache}, G. and {Lamarre}, J.-M. and {Lasenby}, A. and {Lattanzi}, M. and {Lawrence}, C.~R. and {Le Jeune}, M. and {Lemos}, P. and {Lesgourgues}, J. and {Levrier}, F. and {Lewis}, A. and {Liguori}, M. and {Lilje}, P.~B. and {Lilley}, M. and {Lindholm}, V. and {L{\'o}pez-Caniego}, M. and {Lubin}, P.~M. and {Ma}, Y.-Z. and {Mac{\'\i}as-P{\'e}rez}, J.~F. and {Maggio}, G. and {Maino}, D. and {Mandolesi}, N. and {Mangilli}, A. and {Marcos-Caballero}, A. and {Maris}, M. and {Martin}, P.~G. and {Martinelli}, M. and {Mart{\'\i}nez-Gonz{\'a}lez}, E. and {Matarrese}, S. and {Mauri}, N. and {McEwen}, J.~D. and {Meinhold}, P.~R. and {Melchiorri}, A. and {Mennella}, A. and {Migliaccio}, M. and {Millea}, M. and {Mitra}, S. and {Miville-Desch{\^e}nes}, M.-A. and {Molinari}, D. and {Montier}, L. and {Morgante}, G. and {Moss}, A. and {Natoli}, P. and {N{\o}rgaard-Nielsen}, H.~U. and {Pagano}, L. and {Paoletti}, D. and {Partridge}, B. and {Patanchon}, G. and {Peiris}, H.~V. and {Perrotta}, F. and {Pettorino}, V. and {Piacentini}, F. and {Polastri}, L. and {Polenta}, G. and {Puget}, J.-L. and {Rachen}, J.~P. and {Reinecke}, M. and {Remazeilles}, M. and {Renzi}, A. and {Rocha}, G. and {Rosset}, C. and {Roudier}, G. and {Rubi{\~n}o-Mart{\'\i}n}, J.~A. and {Ruiz-Granados}, B. and {Salvati}, L. and {Sandri}, M. and {Savelainen}, M. and {Scott}, D. and {Shellard}, E.~P.~S. and {Sirignano}, C. and {Sirri}, G. and {Spencer}, L.~D. and {Sunyaev}, R. and {Suur-Uski}, A.-S. and {Tauber}, J.~A. and {Tavagnacco}, D. and {Tenti}, M. and {Toffolatti}, L. and {Tomasi}, M. and {Trombetti}, T. and {Valenziano}, L. and {Valiviita}, J. and {Van Tent}, B. and {Vibert}, L. and {Vielva}, P. and {Villa}, F. and {Vittorio}, N. and {Wandelt}, B.~D. and {Wehus}, I.~K. and {White}, M. and {White}, S.~D.~M. and {Zacchei}, A. and {Zonca}, A.},
        title = "{Planck 2018 results. VI. Cosmological parameters}",
      journal = {\aap},
         year = 2020,
        month = sep,
       volume = {641},
          eid = {A6},
        pages = {A6},
          doi = {10.1051/0004-6361/201833910},
archivePrefix = {arXiv},
       eprint = {1807.06209},
 primaryClass = {astro-ph.CO},
       adsurl = {https://ui.adsabs.harvard.edu/abs/2020A&A...641A...6P}
}

@ARTICLE{Valentino2021CQGra,
       author = {{Di Valentino}, Eleonora and {Mena}, Olga and {Pan}, Supriya and {Visinelli}, Luca and {Yang}, Weiqiang and {Melchiorri}, Alessandro and {Mota}, David F. and {Riess}, Adam G. and {Silk}, Joseph},
        title = "{In the realm of the Hubble tension-a review of solutions}",
      journal = {Classical and Quantum Gravity},
         year = 2021,
        month = jul,
       volume = {38},
       number = {15},
          eid = {153001},
        pages = {153001},
          doi = {10.1088/1361-6382/ac086d},
archivePrefix = {arXiv},
       eprint = {2103.01183},
 primaryClass = {astro-ph.CO},
       adsurl = {https://ui.adsabs.harvard.edu/abs/2021CQGra..38o3001D}
}

@ARTICLE{Abbott2019PhRvX,
       author = {{Abbott}, B.~P. and {Abbott}, R. and {Abbott}, T.~D. and {Abraham}, S. and {Acernese}, F. and {Ackley}, K. and {Adams}, C. and {Adhikari}, R.~X. and {Adya}, V.~B. and {Affeldt}, C. and {Agathos}, M. and {Agatsuma}, K. and {Aggarwal}, N. and {Aguiar}, O.~D. and {Aiello}, L. and {Ain}, A. and {Ajith}, P. and {Allen}, G. and {Allocca}, A. and {Aloy}, M.~A. and {Altin}, P.~A. and {Amato}, A. and {Ananyeva}, A. and {Anderson}, S.~B. and {Anderson}, W.~G. and {Angelova}, S.~V. and {Antier}, S. and {Appert}, S. and {Arai}, K. and {Araya}, M.~C. and {Areeda}, J.~S. and {Ar{\`e}ne}, M. and {Arnaud}, N. and {Arun}, K.~G. and {Ascenzi}, S. and {Ashton}, G. and {Aston}, S.~M. and {Astone}, P. and {Aubin}, F. and {Aufmuth}, P. and {AultONeal}, K. and {Austin}, C. and {Avendano}, V. and {Avila-Alvarez}, A. and {Babak}, S. and {Bacon}, P. and {Badaracco}, F. and {Bader}, M.~K.~M. and {Bae}, S. and {Baker}, P.~T. and {Baldaccini}, F. and {Ballardin}, G. and {Ballmer}, S.~W. and {Banagiri}, S. and {Barayoga}, J.~C. and {Barclay}, S.~E. and {Barish}, B.~C. and {Barker}, D. and {Barkett}, K. and {Barnum}, S. and {Barone}, F. and {Barr}, B. and {Barsotti}, L. and {Barsuglia}, M. and {Barta}, D. and {Bartlett}, J. and {Bartos}, I. and {Bassiri}, R. and {Basti}, A. and {Bawaj}, M. and {Bayley}, J.~C. and {Bazzan}, M. and {B{\'e}csy}, B. and {Bejger}, M. and {Belahcene}, I. and {Bell}, A.~S. and {Beniwal}, D. and {Berger}, B.~K. and {Bergmann}, G. and {Bernuzzi}, S. and {Bero}, J.~J. and {Berry}, C.~P.~L. and {Bersanetti}, D. and {Bertolini}, A. and {Betzwieser}, J. and {Bhandare}, R. and {Bidler}, J. and {Bilenko}, I.~A. and {Bilgili}, S.~A. and {Billingsley}, G. and {Birch}, J. and {Birney}, R. and {Birnholtz}, O. and {Biscans}, S. and {Biscoveanu}, S. and {Bisht}, A. and {Bitossi}, M. and {Bizouard}, M.~A. and {Blackburn}, J.~K. and {Blackman}, J. and {Blair}, C.~D. and {Blair}, D.~G. and {Blair}, R.~M. and {Bloemen}, S. and {Bode}, N. and {Boer}, M. and {Boetzel}, Y. and {Bogaert}, G. and {Bondu}, F. and {Bonilla}, E. and {Bonnand}, R. and {Booker}, P. and {Boom}, B.~A. and {Booth}, C.~D. and {Bork}, R. and {Boschi}, V. and {Bose}, S. and {Bossie}, K. and {Bossilkov}, V. and {Bosveld}, J. and {Bouffanais}, Y. and {Bozzi}, A. and {Bradaschia}, C. and {Brady}, P.~R. and {Bramley}, A. and {Branchesi}, M. and {Brau}, J.~E. and {Briant}, T. and {Briggs}, J.~H. and {Brighenti}, F. and {Brillet}, A. and {Brinkmann}, M. and {Brisson}, V. and {Brockill}, P. and {Brooks}, A.~F. and {Brown}, D.~D. and {Brunett}, S. and {Buikema}, A. and {Bulik}, T. and {Bulten}, H.~J. and {Buonanno}, A. and {Buskulic}, D. and {Bustamante Rosell}, M.~J. and {Buy}, C. and {Byer}, R.~L. and {Cabero}, M. and {Cadonati}, L. and {Cagnoli}, G. and {Cahillane}, C. and {Calder{\'o}n Bustillo}, J. and {Callister}, T.~A. and {Calloni}, E. and {Camp}, J.~B. and {Campbell}, W.~A. and {Canepa}, M. and {Cannon}, K.~C. and {Cao}, H. and {Cao}, J. and {Capocasa}, E. and {Carbognani}, F. and {Caride}, S. and {Carney}, M.~F. and {Carullo}, G. and {Casanueva Diaz}, J. and {Casentini}, C. and {Caudill}, S. and {Cavagli{\`a}}, M. and {Cavalier}, F. and {Cavalieri}, R. and {Cella}, G. and {Cerd{\'a}-Dur{\'a}n}, P. and {Cerretani}, G. and {Cesarini}, E. and {Chaibi}, O. and {Chakravarti}, K. and {Chamberlin}, S.~J. and {Chan}, M. and {Chao}, S. and {Charlton}, P. and {Chase}, E.~A. and {Chassande-Mottin}, E. and {Chatterjee}, D. and {Chaturvedi}, M. and {Chatziioannou}, K. and {Cheeseboro}, B.~D. and {Chen}, H.~Y. and {Chen}, X. and {Chen}, Y. and {Cheng}, H.-P. and {Cheong}, C.~K. and {Chia}, H.~Y. and {Chincarini}, A. and {Chiummo}, A. and {Cho}, G. and {Cho}, H.~S. and {Cho}, M. and {Christensen}, N. and {Chu}, Q. and {Chua}, S. and {Chung}, K.~W.},
        title = "{GWTC-1: A Gravitational-Wave Transient Catalog of Compact Binary Mergers Observed by LIGO and Virgo during the First and Second Observing Runs}",
      journal = {Physical Review X},
         year = 2019,
        month = jul,
       volume = {9},
       number = {3},
          eid = {031040},
        pages = {031040},
          doi = {10.1103/PhysRevX.9.031040},
archivePrefix = {arXiv},
       eprint = {1811.12907},
 primaryClass = {astro-ph.HE},
       adsurl = {https://ui.adsabs.harvard.edu/abs/2019PhRvX...9c1040A}
}

@ARTICLE{LIGO2025arXiv,
       author = {{The LIGO Scientific Collaboration} and {the Virgo Collaboration} and {the KAGRA Collaboration} and {Abac}, A.~G. and {Abouelfettouh}, I. and {Acernese}, F. and {Ackley}, K. and {Adamcewicz}, C. and {Adhicary}, S. and {Adhikari}, D. and {Adhikari}, N. and {Adhikari}, R.~X. and {Adkins}, V.~K. and {Afroz}, S. and {Agapito}, A. and {Agarwal}, D. and {Agathos}, M. and {Aggarwal}, N. and {Aggarwal}, S. and {Aguiar}, O.~D. and {Ahrend}, I. -L. and {Aiello}, L. and {Ain}, A. and {Ajith}, P. and {Akutsu}, T. and {Albanesi}, S. and {Ali}, W. and {Al-Kershi}, S. and {All{\'e}n{\'e}}, C. and {Allocca}, A. and {Al-Shammari}, S. and {Altin}, P.~A. and {Alvarez-Lopez}, S. and {Amar}, W. and {Amarasinghe}, O. and {Amato}, A. and {Amicucci}, F. and {Amra}, C. and {Ananyeva}, A. and {Anderson}, S.~B. and {Anderson}, W.~G. and {Andia}, M. and {Ando}, M. and {Andr{\'e}s-Carcasona}, M. and {Andri{\'c}}, T. and {Anglin}, J. and {Ansoldi}, S. and {Antelis}, J.~M. and {Antier}, S. and {Aoumi}, M. and {Appavuravther}, E.~Z. and {Appert}, S. and {Apple}, S.~K. and {Arai}, K. and {Araya}, A. and {Araya}, M.~C. and {Arca Sedda}, M. and {Areeda}, J.~S. and {Aritomi}, N. and {Armato}, F. and {Armstrong}, S. and {Arnaud}, N. and {Arogeti}, M. and {Aronson}, S.~M. and {Arun}, K.~G. and {Ashton}, G. and {Aso}, Y. and {Asprea}, L. and {Assiduo}, M. and {Assis de Souza Melo}, S. and {Aston}, S.~M. and {Astone}, P. and {Attadio}, F. and {Aubin}, F. and {AultONeal}, K. and {Avallone}, G. and {Avila}, E.~A. and {Babak}, S. and {Badger}, C. and {Bae}, S. and {Bagnasco}, S. and {Baiotti}, L. and {Bajpai}, R. and {Baka}, T. and {Baker}, A.~M. and {Baker}, K.~A. and {Baker}, T. and {Baldi}, G. and {Baldicchi}, N. and {Ball}, M. and {Ballardin}, G. and {Ballmer}, S.~W. and {Banagiri}, S. and {Banerjee}, B. and {Bankar}, D. and {Baptiste}, T.~M. and {Baral}, P. and {Baratti}, M. and {Barayoga}, J.~C. and {Barish}, B.~C. and {Barker}, D. and {Barman}, N. and {Barneo}, P. and {Barone}, F. and {Barr}, B. and {Barsotti}, L. and {Barsuglia}, M. and {Barta}, D. and {Bartoletti}, A.~M. and {Barton}, M.~A. and {Bartos}, I. and {Basalaev}, A. and {Bassiri}, R. and {Basti}, A. and {Bawaj}, M. and {Baxi}, P. and {Bayley}, J.~C. and {Baylor}, A.~C. and {Baynard}, II, P.~A. and {Bazzan}, M. and {Bedakihale}, V.~M. and {Beirnaert}, F. and {Bejger}, M. and {Belardinelli}, D. and {Bell}, A.~S. and {Bellie}, D.~S. and {Bellizzi}, L. and {Benoit}, W. and {Bentara}, I. and {Bentley}, J.~D. and {Ben Yaala}, M. and {Bera}, S. and {Bergamin}, F. and {Berger}, B.~K. and {Bernuzzi}, S. and {Beroiz}, M. and {Berry}, C.~P.~L. and {Bersanetti}, D. and {Bertheas}, T. and {Bertolini}, A. and {Betzwieser}, J. and {Beveridge}, D. and {Bevilacqua}, G. and {Bevins}, N. and {Bhandare}, R. and {Bhatt}, R. and {Bhattacharjee}, D. and {Bhattacharyya}, S. and {Bhaumik}, S. and {Biancalana}, V. and {Bianchi}, A. and {Bilenko}, I.~A. and {Billingsley}, G. and {Binetti}, A. and {Bini}, S. and {Binu}, C. and {Biot}, S. and {Birnholtz}, O. and {Biscoveanu}, S. and {Bisht}, A. and {Bitossi}, M. and {Bizouard}, M. -A. and {Blaber}, S. and {Blackburn}, J.~K. and {Blagg}, L.~A. and {Blair}, C.~D. and {Blair}, D.~G. and {Bode}, N. and {Boettner}, N. and {Boileau}, G. and {Boldrini}, M. and {Bolingbroke}, G.~N. and {Bolliand}, A. and {Bonavena}, L.~D. and {Bondarescu}, R. and {Bondu}, F. and {Bonilla}, E. and {Bonilla}, M.~S. and {Bonino}, A. and {Bonnand}, R. and {Borchers}, A. and {Borhanian}, S. and {Boschi}, V. and {Bose}, S. and {Bossilkov}, V. and {Bothra}, Y. and {Boudon}, A. and {Bourg}, L. and {Boyle}, M. and {Bozzi}, A. and {Bradaschia}, C. and {Brady}, P.~R. and {Branch}, A. and {Branchesi}, M. and {Braun}, I. and {Briant}, T. and {Brillet}, A. and {Brinkmann}, M. and {Brockill}, P. and {Brockmueller}, E.},
        title = "{GWTC-4.0: Updating the Gravitational-Wave Transient Catalog with Observations from the First Part of the Fourth LIGO-Virgo-KAGRA Observing Run}",
      journal = {arXiv e-prints},
         year = 2025,
        month = aug,
          eid = {arXiv:2508.18082},
        pages = {arXiv:2508.18082},
          doi = {10.48550/arXiv.2508.18082},
archivePrefix = {arXiv},
       eprint = {2508.18082},
 primaryClass = {gr-qc},
       adsurl = {https://ui.adsabs.harvard.edu/abs/2025arXiv250818082T}
}

@ARTICLE{Abbott2024PhRvD,
       author = {{Abbott}, R. and {Abbott}, T.~D. and {Acernese}, F. and {Ackley}, K. and {Adams}, C. and {Adhikari}, N. and {Adhikari}, R.~X. and {Adya}, V.~B. and {Affeldt}, C. and {Agarwal}, D. and {Agathos}, M. and {Agatsuma}, K. and {Aggarwal}, N. and {Aguiar}, O.~D. and {Aiello}, L. and {Ain}, A. and {Ajith}, P. and {Albanesi}, S. and {Allocca}, A. and {Altin}, P.~A. and {Amato}, A. and {Anand}, C. and {Anand}, S. and {Ananyeva}, A. and {Anderson}, S.~B. and {Anderson}, W.~G. and {Andrade}, T. and {Andres}, N. and {Andri{\'c}}, T. and {Angelova}, S.~V. and {Ansoldi}, S. and {Antelis}, J.~M. and {Antier}, S. and {Appert}, S. and {Arai}, K. and {Araya}, M.~C. and {Areeda}, J.~S. and {Ar{\`e}ne}, M. and {Arnaud}, N. and {Aronson}, S.~M. and {Arun}, K.~G. and {Asali}, Y. and {Ashton}, G. and {Assiduo}, M. and {Aston}, S.~M. and {Astone}, P. and {Aubin}, F. and {Austin}, C. and {Babak}, S. and {Badaracco}, F. and {Bader}, M.~K.~M. and {Badger}, C. and {Bae}, S. and {Baer}, A.~M. and {Bagnasco}, S. and {Bai}, Y. and {Baird}, J. and {Ball}, M. and {Ballardin}, G. and {Ballmer}, S.~W. and {Balsamo}, A. and {Baltus}, G. and {Banagiri}, S. and {Bankar}, D. and {Barayoga}, J.~C. and {Barbieri}, C. and {Barish}, B.~C. and {Barker}, D. and {Barneo}, P. and {Barone}, F. and {Barr}, B. and {Barsotti}, L. and {Barsuglia}, M. and {Barta}, D. and {Bartlett}, J. and {Barton}, M.~A. and {Bartos}, I. and {Bassiri}, R. and {Basti}, A. and {Bawaj}, M. and {Bayley}, J.~C. and {Baylor}, A.~C. and {Bazzan}, M. and {B{\'e}csy}, B. and {Bedakihale}, V.~M. and {Bejger}, M. and {Belahcene}, I. and {Benedetto}, V. and {Beniwal}, D. and {Bennett}, T.~F. and {Bentley}, J.~D. and {BenYaala}, M. and {Bergamin}, F. and {Berger}, B.~K. and {Bernuzzi}, S. and {Berry}, C.~P.~L. and {Bersanetti}, D. and {Bertolini}, A. and {Betzwieser}, J. and {Beveridge}, D. and {Bhandare}, R. and {Bhardwaj}, U. and {Bhattacharjee}, D. and {Bhaumik}, S. and {Bilenko}, I.~A. and {Billingsley}, G. and {Bini}, S. and {Birney}, R. and {Birnholtz}, O. and {Biscans}, S. and {Bischi}, M. and {Biscoveanu}, S. and {Bisht}, A. and {Biswas}, B. and {Bitossi}, M. and {Bizouard}, M.-A. and {Blackburn}, J.~K. and {Blair}, C.~D. and {Blair}, D.~G. and {Blair}, R.~M. and {Bobba}, F. and {Bode}, N. and {Boer}, M. and {Bogaert}, G. and {Boldrini}, M. and {Bonavena}, L.~D. and {Bondu}, F. and {Bonilla}, E. and {Bonnand}, R. and {Booker}, P. and {Boom}, B.~A. and {Bork}, R. and {Boschi}, V. and {Bose}, N. and {Bose}, S. and {Bossilkov}, V. and {Boudart}, V. and {Bouffanais}, Y. and {Bozzi}, A. and {Bradaschia}, C. and {Brady}, P.~R. and {Bramley}, A. and {Branch}, A. and {Branchesi}, M. and {Brau}, J.~E. and {Breschi}, M. and {Briant}, T. and {Briggs}, J.~H. and {Brillet}, A. and {Brinkmann}, M. and {Brockill}, P. and {Brooks}, A.~F. and {Brooks}, J. and {Brown}, D.~D. and {Brunett}, S. and {Bruno}, G. and {Bruntz}, R. and {Bryant}, J. and {Bulik}, T. and {Bulten}, H.~J. and {Buonanno}, A. and {Buscicchio}, R. and {Buskulic}, D. and {Buy}, C. and {Byer}, R.~L. and {Cadonati}, L. and {Cagnoli}, G. and {Cahillane}, C. and {Bustillo}, J. Calder{\'o}n and {Callaghan}, J.~D. and {Callister}, T.~A. and {Calloni}, E. and {Cameron}, J. and {Camp}, J.~B. and {Canepa}, M. and {Canevarolo}, S. and {Cannavacciuolo}, M. and {Cannon}, K.~C. and {Cao}, H. and {Capote}, E. and {Carapella}, G. and {Carbognani}, F. and {Carlin}, J.~B. and {Carney}, M.~F. and {Carpinelli}, M. and {Carrillo}, G. and {Carullo}, G. and {Carver}, T.~L. and {Diaz}, J. Casanueva and {Casentini}, C. and {Castaldi}, G. and {Caudill}, S. and {Cavagli{\`a}}, M. and {Cavalier}, F. and {Cavalieri}, R. and {Ceasar}, M. and {Cella}, G. and {Cerd{\'a}-Dur{\'a}n}, P. and {Cesarini}, E. and {Chaibi}, W.},
        title = "{GWTC-2.1: Deep extended catalog of compact binary coalescences observed by LIGO and Virgo during the first half of the third observing run}",
      journal = {\prd},
         year = 2024,
        month = jan,
       volume = {109},
       number = {2},
          eid = {022001},
        pages = {022001},
          doi = {10.1103/PhysRevD.109.022001},
archivePrefix = {arXiv},
       eprint = {2108.01045},
 primaryClass = {gr-qc},
       adsurl = {https://ui.adsabs.harvard.edu/abs/2024PhRvD.109b2001A}
}

@ARTICLE{Abbott2023PhRvX,
       author = {{Abbott}, R. and {Abbott}, T.~D. and {Acernese}, F. and {Ackley}, K. and {Adams}, C. and {Adhikari}, N. and {Adhikari}, R.~X. and {Adya}, V.~B. and {Affeldt}, C. and {Agarwal}, D. and {Agathos}, M. and {Agatsuma}, K. and {Aggarwal}, N. and {Aguiar}, O.~D. and {Aiello}, L. and {Ain}, A. and {Ajith}, P. and {Akcay}, S. and {Akutsu}, T. and {Albanesi}, S. and {Allocca}, A. and {Altin}, P.~A. and {Amato}, A. and {Anand}, C. and {Anand}, S. and {Ananyeva}, A. and {Anderson}, S.~B. and {Anderson}, W.~G. and {Ando}, M. and {Andrade}, T. and {Andres}, N. and {Andri{\'c}}, T. and {Angelova}, S.~V. and {Ansoldi}, S. and {Antelis}, J.~M. and {Antier}, S. and {Appert}, S. and {Arai}, Koji and {Arai}, Koya and {Arai}, Y. and {Araki}, S. and {Araya}, A. and {Araya}, M.~C. and {Areeda}, J.~S. and {Ar{\`e}ne}, M. and {Aritomi}, N. and {Arnaud}, N. and {Arogeti}, M. and {Aronson}, S.~M. and {Arun}, K.~G. and {Asada}, H. and {Asali}, Y. and {Ashton}, G. and {Aso}, Y. and {Assiduo}, M. and {Aston}, S.~M. and {Astone}, P. and {Aubin}, F. and {Austin}, C. and {Babak}, S. and {Badaracco}, F. and {Bader}, M.~K.~M. and {Badger}, C. and {Bae}, S. and {Bae}, Y. and {Baer}, A.~M. and {Bagnasco}, S. and {Bai}, Y. and {Baiotti}, L. and {Baird}, J. and {Bajpai}, R. and {Ball}, M. and {Ballardin}, G. and {Ballmer}, S.~W. and {Balsamo}, A. and {Baltus}, G. and {Banagiri}, S. and {Bankar}, D. and {Barayoga}, J.~C. and {Barbieri}, C. and {Barish}, B.~C. and {Barker}, D. and {Barneo}, P. and {Barone}, F. and {Barr}, B. and {Barsotti}, L. and {Barsuglia}, M. and {Barta}, D. and {Bartlett}, J. and {Barton}, M.~A. and {Bartos}, I. and {Bassiri}, R. and {Basti}, A. and {Bawaj}, M. and {Bayley}, J.~C. and {Baylor}, A.~C. and {Bazzan}, M. and {B{\'e}csy}, B. and {Bedakihale}, V.~M. and {Bejger}, M. and {Belahcene}, I. and {Benedetto}, V. and {Beniwal}, D. and {Bennett}, T.~F. and {Bentley}, J.~D. and {Benyaala}, M. and {Bergamin}, F. and {Berger}, B.~K. and {Bernuzzi}, S. and {Berry}, C.~P.~L. and {Bersanetti}, D. and {Bertolini}, A. and {Betzwieser}, J. and {Beveridge}, D. and {Bhandare}, R. and {Bhardwaj}, U. and {Bhattacharjee}, D. and {Bhaumik}, S. and {Bilenko}, I.~A. and {Billingsley}, G. and {Bini}, S. and {Birney}, R. and {Birnholtz}, O. and {Biscans}, S. and {Bischi}, M. and {Biscoveanu}, S. and {Bisht}, A. and {Biswas}, B. and {Bitossi}, M. and {Bizouard}, M.-A. and {Blackburn}, J.~K. and {Blair}, C.~D. and {Blair}, D.~G. and {Blair}, R.~M. and {Bobba}, F. and {Bode}, N. and {Boer}, M. and {Bogaert}, G. and {Boldrini}, M. and {Bonavena}, L.~D. and {Bondu}, F. and {Bonilla}, E. and {Bonnand}, R. and {Booker}, P. and {Boom}, B.~A. and {Bork}, R. and {Boschi}, V. and {Bose}, N. and {Bose}, S. and {Bossilkov}, V. and {Boudart}, V. and {Bouffanais}, Y. and {Bozzi}, A. and {Bradaschia}, C. and {Brady}, P.~R. and {Bramley}, A. and {Branch}, A. and {Branchesi}, M. and {Brandt}, J. and {Brau}, J.~E. and {Breschi}, M. and {Briant}, T. and {Briggs}, J.~H. and {Brillet}, A. and {Brinkmann}, M. and {Brockill}, P. and {Brooks}, A.~F. and {Brooks}, J. and {Brown}, D.~D. and {Brunett}, S. and {Bruno}, G. and {Bruntz}, R. and {Bryant}, J. and {Bulik}, T. and {Bulten}, H.~J. and {Buonanno}, A. and {Buscicchio}, R. and {Buskulic}, D. and {Buy}, C. and {Byer}, R.~L. and {Davies}, G.~S. Cabourn and {Cadonati}, L. and {Cagnoli}, G. and {Cahillane}, C. and {Bustillo}, J. Calder{\'o}n and {Callaghan}, J.~D. and {Callister}, T.~A. and {Calloni}, E. and {Cameron}, J. and {Camp}, J.~B. and {Canepa}, M. and {Canevarolo}, S. and {Cannavacciuolo}, M. and {Cannon}, K.~C. and {Cao}, H. and {Cao}, Z. and {Capocasa}, E. and {Capote}, E. and {Carapella}, G. and {Carbognani}, F.},
        title = "{GWTC--3: Compact Binary Coalescences Observed by LIGO and Virgo during the Second Part of the Third Observing Run}",
      journal = {Physical Review X},
         year = 2023,
        month = oct,
       volume = {13},
       number = {4},
          eid = {041039},
        pages = {041039},
          doi = {10.1103/PhysRevX.13.041039},
archivePrefix = {arXiv},
       eprint = {2111.03606},
 primaryClass = {gr-qc},
       adsurl = {https://ui.adsabs.harvard.edu/abs/2023PhRvX..13d1039A}
}

@ARTICLE{Abbott2023ApJ,
       author = {{Abbott}, R. and {Abe}, H. and {Acernese}, F. and {Ackley}, K. and {Adhikari}, N. and {Adhikari}, R.~X. and {Adkins}, V.~K. and {Adya}, V.~B. and {Affeldt}, C. and {Agarwal}, D. and {Agathos}, M. and {Agatsuma}, K. and {Aggarwal}, N. and {Aguiar}, O.~D. and {Aiello}, L. and {Ain}, A. and {Ajith}, P. and {Akutsu}, T. and {Albanesi}, S. and {Alfaidi}, R.~A. and {Allocca}, A. and {Altin}, P.~A. and {Amato}, A. and {Anand}, C. and {Anand}, S. and {Ananyeva}, A. and {Anderson}, S.~B. and {Anderson}, W.~G. and {Ando}, M. and {Andrade}, T. and {Andres}, N. and {Andr{\'e}s-Carcasona}, M. and {Andri{\'c}}, T. and {Angelova}, S.~V. and {Ansoldi}, S. and {Antelis}, J.~M. and {Antier}, S. and {Apostolatos}, T. and {Appavuravther}, E.~Z. and {Appert}, S. and {Apple}, S.~K. and {Arai}, K. and {Araya}, A. and {Araya}, M.~C. and {Areeda}, J.~S. and {Ar{\`e}ne}, M. and {Aritomi}, N. and {Arnaud}, N. and {Arogeti}, M. and {Aronson}, S.~M. and {Arun}, K.~G. and {Asada}, H. and {Asali}, Y. and {Ashton}, G. and {Aso}, Y. and {Assiduo}, M. and {de Souza Melo}, S. Assis and {Aston}, S.~M. and {Astone}, P. and {Aubin}, F. and {Aultoneal}, K. and {Austin}, C. and {Babak}, S. and {Badaracco}, F. and {Bader}, M.~K.~M. and {Badger}, C. and {Bae}, S. and {Bae}, Y. and {Baer}, A.~M. and {Bagnasco}, S. and {Bai}, Y. and {Baird}, J. and {Bajpai}, R. and {Baka}, T. and {Ball}, M. and {Ballardin}, G. and {Ballmer}, S.~W. and {Balsamo}, A. and {Baltus}, G. and {Banagiri}, S. and {Banerjee}, B. and {Bankar}, D. and {Barayoga}, J.~C. and {Barbieri}, C. and {Barbieri}, R. and {Barish}, B.~C. and {Barker}, D. and {Barneo}, P. and {Barone}, F. and {Barr}, B. and {Barsotti}, L. and {Barsuglia}, M. and {Barta}, D. and {Bartlett}, J. and {Barton}, M.~A. and {Bartos}, I. and {Basak}, S. and {Bassiri}, R. and {Basti}, A. and {Bawaj}, M. and {Bayley}, J.~C. and {Bazzan}, M. and {Becher}, B.~R. and {B{\'e}csy}, B. and {Bedakihale}, V.~M. and {Beirnaert}, F. and {Bejger}, M. and {Belahcene}, I. and {Benedetto}, V. and {Beniwal}, D. and {Benjamin}, M.~G. and {Bennett}, T.~F. and {Bentley}, J.~D. and {Benyaala}, M. and {Bera}, S. and {Berbel}, M. and {Bergamin}, F. and {Berger}, B.~K. and {Bernuzzi}, S. and {Berry}, C.~P.~L. and {Bersanetti}, D. and {Bertolini}, A. and {Betzwieser}, J. and {Beveridge}, D. and {Bhandare}, R. and {Bhandari}, A.~V. and {Bhardwaj}, U. and {Bhatt}, R. and {Bhattacharjee}, D. and {Bhaumik}, S. and {Bianchi}, A. and {Bilenko}, I.~A. and {Billingsley}, G. and {Bilicki}, M. and {Bini}, S. and {Birney}, I.~A. and {Birnholtz}, O. and {Biscans}, S. and {Bischi}, M. and {Biscoveanu}, S. and {Bisht}, A. and {Biswas}, B. and {Bitossi}, M. and {Bizouard}, M.-A. and {Blackburn}, J.~K. and {Blair}, C.~D. and {Blair}, D.~G. and {Blair}, R.~M. and {Bobba}, F. and {Bode}, N. and {Bo{\"e}r}, M. and {Bogaert}, G. and {Boldrini}, M. and {Bolingbroke}, G.~N. and {Bonavena}, L.~D. and {Bondu}, F. and {Bonilla}, E. and {Bonnand}, R. and {Booker}, P. and {Boom}, B.~A. and {Bork}, R. and {Boschi}, V. and {Bose}, N. and {Bose}, S. and {Bossilkov}, V. and {Boudart}, V. and {Bouffanais}, Y. and {Bozzi}, A. and {Bradaschia}, C. and {Brady}, P.~R. and {Bramley}, A. and {Branch}, A. and {Branchesi}, M. and {Brau}, J.~E. and {Breschi}, M. and {Briant}, T. and {Briggs}, J.~H. and {Brillet}, A. and {Brinkmann}, M. and {Brockill}, P. and {Brooks}, A.~F. and {Brooks}, J. and {Brown}, D.~D. and {Brunett}, S. and {Bruno}, G. and {Bruntz}, R. and {Bryant}, J. and {Bucci}, F. and {Bulik}, T. and {Bulten}, H.~J. and {Buonanno}, A. and {Burtnyk}, K. and {Buscicchio}, R. and {Buskulic}, D. and {Buy}, C. and {Byer}, R.~L. and {Davies}, G.~S. Cabourn and {Cabras}, G. and {Cabrita}, R. and {Cadonati}, L.},
        title = "{Constraints on the Cosmic Expansion History from GWTC--3}",
      journal = {\apj},
         year = 2023,
        month = jun,
       volume = {949},
       number = {2},
          eid = {76},
        pages = {76},
          doi = {10.3847/1538-4357/ac74bb},
archivePrefix = {arXiv},
       eprint = {2111.03604},
 primaryClass = {astro-ph.CO},
       adsurl = {https://ui.adsabs.harvard.edu/abs/2023ApJ...949...76A}
}

@ARTICLE{Abbott2021PhRvX,
       author = {{Abbott}, R. and {Abbott}, T.~D. and {Abraham}, S. and {Acernese}, F. and {Ackley}, K. and {Adams}, A. and {Adams}, C. and {Adhikari}, R.~X. and {Adya}, V.~B. and {Affeldt}, C. and {Agathos}, M. and {Agatsuma}, K. and {Aggarwal}, N. and {Aguiar}, O.~D. and {Aiello}, L. and {Ain}, A. and {Ajith}, P. and {Akcay}, S. and {Allen}, G. and {Allocca}, A. and {Altin}, P.~A. and {Amato}, A. and {Anand}, S. and {Ananyeva}, A. and {Anderson}, S.~B. and {Anderson}, W.~G. and {Angelova}, S.~V. and {Ansoldi}, S. and {Antelis}, J.~M. and {Antier}, S. and {Appert}, S. and {Arai}, K. and {Araya}, M.~C. and {Areeda}, J.~S. and {Ar{\`e}ne}, M. and {Arnaud}, N. and {Aronson}, S.~M. and {Arun}, K.~G. and {Asali}, Y. and {Ascenzi}, S. and {Ashton}, G. and {Aston}, S.~M. and {Astone}, P. and {Aubin}, F. and {Aufmuth}, P. and {AultONeal}, K. and {Austin}, C. and {Avendano}, V. and {Babak}, S. and {Badaracco}, F. and {Bader}, M.~K.~M. and {Bae}, S. and {Baer}, A.~M. and {Bagnasco}, S. and {Baird}, J. and {Ball}, M. and {Ballardin}, G. and {Ballmer}, S.~W. and {Bals}, A. and {Balsamo}, A. and {Baltus}, G. and {Banagiri}, S. and {Bankar}, D. and {Bankar}, R.~S. and {Barayoga}, J.~C. and {Barbieri}, C. and {Barish}, B.~C. and {Barker}, D. and {Barneo}, P. and {Barnum}, S. and {Barone}, F. and {Barr}, B. and {Barsotti}, L. and {Barsuglia}, M. and {Barta}, D. and {Bartlett}, J. and {Bartos}, I. and {Bassiri}, R. and {Basti}, A. and {Bawaj}, M. and {Bayley}, J.~C. and {Bazzan}, M. and {Becher}, B.~R. and {B{\'e}csy}, B. and {Bedakihale}, V.~M. and {Bejger}, M. and {Belahcene}, I. and {Beniwal}, D. and {Benjamin}, M.~G. and {Bennett}, T.~F. and {Bentley}, J.~D. and {Bergamin}, F. and {Berger}, B.~K. and {Bergmann}, G. and {Bernuzzi}, S. and {Berry}, C.~P.~L. and {Bersanetti}, D. and {Bertolini}, A. and {Betzwieser}, J. and {Bhandare}, R. and {Bhandari}, A.~V. and {Bhattacharjee}, D. and {Bidler}, J. and {Bilenko}, I.~A. and {Billingsley}, G. and {Birney}, R. and {Birnholtz}, O. and {Biscans}, S. and {Bischi}, M. and {Biscoveanu}, S. and {Bisht}, A. and {Bitossi}, M. and {Bizouard}, M.-A. and {Blackburn}, J.~K. and {Blackman}, J. and {Blair}, C.~D. and {Blair}, D.~G. and {Blair}, R.~M. and {Blanch}, O. and {Bobba}, F. and {Bode}, N. and {Boer}, M. and {Boetzel}, Y. and {Bogaert}, G. and {Boldrini}, M. and {Bondu}, F. and {Bonilla}, E. and {Bonnand}, R. and {Booker}, P. and {Boom}, B.~A. and {Bork}, R. and {Boschi}, V. and {Bose}, S. and {Bossilkov}, V. and {Boudart}, V. and {Bouffanais}, Y. and {Bozzi}, A. and {Bradaschia}, C. and {Brady}, P.~R. and {Bramley}, A. and {Branchesi}, M. and {Brau}, J.~E. and {Breschi}, M. and {Briant}, T. and {Briggs}, J.~H. and {Brighenti}, F. and {Brillet}, A. and {Brinkmann}, M. and {Brockill}, P. and {Brooks}, A.~F. and {Brooks}, J. and {Brown}, D.~D. and {Brunett}, S. and {Bruno}, G. and {Bruntz}, R. and {Buikema}, A. and {Bulik}, T. and {Bulten}, H.~J. and {Buonanno}, A. and {Buscicchio}, R. and {Buskulic}, D. and {Byer}, R.~L. and {Cabero}, M. and {Cadonati}, L. and {Caesar}, M. and {Cagnoli}, G. and {Cahillane}, C. and {Calder{\'o}n Bustillo}, J. and {Callaghan}, J.~D. and {Callister}, T.~A. and {Calloni}, E. and {Camp}, J.~B. and {Canepa}, M. and {Cannon}, K.~C. and {Cao}, H. and {Cao}, J. and {Carapella}, G. and {Carbognani}, F. and {Carney}, M.~F. and {Carpinelli}, M. and {Carullo}, G. and {Carver}, T.~L. and {Casanueva Diaz}, J. and {Casentini}, C. and {Caudill}, S. and {Cavagli{\`a}}, M. and {Cavalier}, F. and {Cavalieri}, R. and {Cella}, G. and {Cerd{\'a}-Dur{\'a}n}, P. and {Cesarini}, E. and {Chaibi}, W. and {Chakravarti}, K. and {Chan}, C.-L. and {Chan}, C. and {Chandra}, K. and {Chanial}, P. and {Chao}, S. and {Charlton}, P. and {Chase}, E.~A.},
        title = "{GWTC--2: Compact Binary Coalescences Observed by LIGO and Virgo during the First Half of the Third Observing Run}",
      journal = {Physical Review X},
         year = 2021,
        month = apr,
       volume = {11},
       number = {2},
          eid = {021053},
        pages = {021053},
          doi = {10.1103/PhysRevX.11.021053},
archivePrefix = {arXiv},
       eprint = {2010.14527},
 primaryClass = {gr-qc},
       adsurl = {https://ui.adsabs.harvard.edu/abs/2021PhRvX..11b1053A}
}

@article{Ohanian1974,
  author    = {Hans C. Ohanian},
  title     = {On the focusing of gravitational radiation},
  journal   = {International Journal of Theoretical Physics},
  volume    = {9},
  number    = {6},
  pages     = {425--437},
  year      = {1974},
  month     = {June},
  doi       = {10.1007/BF01810927},
  url       = {https://doi.org/10.1007/BF01810927},
  issn      = {1572-9575}
}

@ARTICLE{Deguchi1986PhRvD,
       author = {{Deguchi}, Shuji and {Watson}, William D.},
        title = "{Wave effects in gravitational lensing of electromagnetic radiation}",
      journal = {\prd},
         year = 1986,
        month = sep,
       volume = {34},
       number = {6},
        pages = {1708-1718},
          doi = {10.1103/PhysRevD.34.1708},
       adsurl = {https://ui.adsabs.harvard.edu/abs/1986PhRvD..34.1708D}
}

@ARTICLE{Wang1996PhRvL,
       author = {{Wang}, Yun and {Stebbins}, Albert and {Turner}, Edwin L.},
        title = "{Gravitational Lensing of Gravitational Waves from Merging Neutron Star Binaries}",
      journal = {\prl},
         year = 1996,
        month = sep,
       volume = {77},
       number = {14},
        pages = {2875-2878},
          doi = {10.1103/PhysRevLett.77.2875},
archivePrefix = {arXiv},
       eprint = {astro-ph/9605140},
 primaryClass = {astro-ph},
       adsurl = {https://ui.adsabs.harvard.edu/abs/1996PhRvL..77.2875W}
}

@ARTICLE{Nakamura1998PhRvL,
       author = {{Nakamura}, Takahiro T.},
        title = "{Gravitational Lensing of Gravitational Waves from Inspiraling Binaries by a Point Mass Lens}",
      journal = {\prl},
         year = 1998,
        month = feb,
       volume = {80},
       number = {6},
        pages = {1138-1141},
          doi = {10.1103/PhysRevLett.80.1138},
       adsurl = {https://ui.adsabs.harvard.edu/abs/1998PhRvL..80.1138N}
}

@ARTICLE{Refsdal1964MNRAS,
       author = {{Refsdal}, S.},
        title = "{On the possibility of determining Hubble's parameter and the masses of galaxies from the gravitational lens effect}",
      journal = {\mnras},
         year = 1964,
        month = jan,
       volume = {128},
        pages = {307},
          doi = {10.1093/mnras/128.4.307},
       adsurl = {https://ui.adsabs.harvard.edu/abs/1964MNRAS.128..307R}
}

%% or
%% [B] Manual formatting (see below)
%% (i) We suggest to always provide author, title and journal data or doi:
%% in short all the informations that clearly identify a document.
%% (ii) please avoid comments such as "For a review'', "For some examples",
%% "and references therein" or move them in the text. In general, please leave only references in the bibliography and move all
%% accessory text in footnotes.
%% (iii) Also, please have only one work for each \bibitem.

% \begin{thebibliography}{99}

% \bibitem{a}
% Author,
% \emph{Title},
% \emph{J. Abbrev.} {\bf vol} (year) pg.

% \bibitem{b}
% Author,
% \emph{Title},
% arxiv:1234.5678.

% \bibitem{c}
% Author,
% \emph{Title},
% Publisher (year).

% \end{thebibliography}
\end{document}